%% file: main-v4.tex
\documentclass[10pt]{ieeetran}

\usepackage{cite}
\usepackage{amsmath,amssymb,amsfonts}
\usepackage{graphicx}
\usepackage{textcomp}

\usepackage[small]{caption}
\usepackage{subcaption}
\usepackage{latexsym}
\usepackage{threeparttable}
\usepackage{threeparttablex}
\usepackage{mathtools}
\usepackage{slashbox}
\usepackage{verbatim}
\usepackage{subfloat}
\usepackage{tabu}
\usepackage{booktabs}
\usepackage{float}
\usepackage{url}
\usepackage{multirow}
\usepackage{graphicx}
\usepackage{cite}
\usepackage{footnote}
\usepackage{array}
\usepackage[table]{xcolor}
\usepackage{tabularx}
\usepackage{array}
\usepackage{slashbox}

\usepackage{algpseudocode}
\usepackage{algorithm}
\usepackage{bm}

\usepackage[letterpaper, top=0.75in, bottom=1.05in, left=0.6in, right=0.6in]{geometry}

\newcolumntype{P}[1]{>{\centering\arraybackslash}p{#1}}

\usepackage{mathtools, nccmath}
\usepackage{mathrsfs}

\usepackage{epstopdf}					
\usepackage{epstopdf}					
\usepackage{amsthm} 

\newtheorem{theorem}{Theorem} 
\newtheorem{lemma}{Lemma} 
\allowdisplaybreaks

\title{Distance-Domain Degrees of Freedom in Near-Field Region}

\author{
    Son T. Duong\IEEEauthorrefmark{1}, 
    Tho Le-Ngoc\IEEEauthorrefmark{1}, \\
    \IEEEauthorblockA{\IEEEauthorrefmark{1}Department of Electrical and Computer Engineering, McGill University, Montreal, Canada} \\
    \IEEEauthorblockA{Email: son.duong@mail.mcgill.ca, tho.le-ngoc@mcgill.ca}
}

\begin{document}

\maketitle

\begin{abstract}
Extremely large aperture array operating in the near‐field regime unlock additional spatial resources, which can be exploited to simultaneously serve multiple users even when they share the same angular direction. This work investigate the distance-domain degrees of freedom (DoF), which are the DoF when the user only varies its distance to the base station and not the angle.
\textcolor{black}{To obtain the distance-domain DoF, we investigate a line-of-sight (LoS) channel between a base station (source) and observation region representing users. The base station is modeled as a large two-dimensional transmit (Tx) array with an arbitrary shape. The observation region is modeled as an arbitrarily long linear receive (Rx) array, where elements are collinearly aligned but located at varying distances from the Tx array.}
We assume that both the Tx and Rx array has continuous aperture with an infinite number of elements and infinitesimal spacing, which establishes an upper bound for the distance-domain DoF in the case of finite number of elements. First, we assume an ideal case where the Tx array is a single piece and the Rx array is on the broad side of the Tx array. By reformulating the channel as an integral operator with a Hermitian convolution kernel, we derive a closed-form expression for the distance-domain DoF via the Fourier transform. Our analysis shows that the distance-domain DoF is predominantly determined by the extreme boundaries of both the Tx and Rx arrays rather than its detailed interior structure. We further extend the framework to non-broadside configurations by employing a projection method that converts the problem to an equivalent broadside case. Finally, we extend our analytical framework to the modular array, which shows the distance-domain DoF gain over the single-piece array given the constraint of the physical length of the array.

\end{abstract}



\input{sec1_introduction_v4}
\input{sec2_1_channelModel_v4}
\input{sec2_2_DoF_definition_v4}

\input{sec3_1_framework_v4}

\input{sec3_2_Fourier_analysis_v4}

\input{sec4_nonBroadside_v4}
\input{sec5_modular_array_v4}

\input{discussion}

\section{Conclusion}
\label{sec:conclusion}
In this paper, we investigated the distance-domain DoF by considering a MIMO channel where the source (represents base station) equips with an Tx array with arbitrary shape and the observation region (represents users) is Rx linear array whose elements are aligned at the same angle but lie at different distances from the Tx array. In an ideal scenario where the Tx array is single-piece and the Rx array is on the broadside of the Tx array, we consider both the arrays having continuous aperture with infinite elements and infinitesimal spacing and model the channel as an integral operator with a non-convolution kernel. We further transform it into the integral operator with a Hermitian convolution kernel, which enables eigenvalue analysis via the Fourier transform. 
\textcolor{black}{This formulation allowed us to derive a closed-form expression for the distance-domain DoF. Our results demonstrate that, unlike the angular-domain DoF, the distance-domain DoF does not scale linearly with the aperture's area. Instead, it is primarily determined by the extreme radial boundaries of the Tx and Rx arrays, regardless of their detailed interior shapes.}

We further extended our analytical framework to more general settings. In the case where Rx array is in the non-broadside of Tx array, we employed a projection method based on Fresnel approximation to transform the non-broadside configurations to an equivalent broadside case. We also evaluate the effect of the error term of the Fresnel approximation, which shows only small indifference between the analytical DoF based on the Fresnel approximation and the simulated DoF based on the exact channel, even when the error term is non-negligible. Moreover, we generalized our analytical framework to the modular array where the Tx array consists of multiple of non-overlapping sub-arrays and compare with the single-piece array. Our analysis also indicates that the central region of the single-piece array contributes less to the distance-domain DoF, which motivates the use of the modular Tx array with a central gap.

\input{appendix_v4}

\bibliographystyle{IEEEtran}
\bibliography{IEEEabrv, references}
\bstctlcite{IEEEexample:BSTcontrol}


\end{document}

%% file: sec1_introduction_v4.tex
\section{Introduction}

The drive toward sixth-generation (6G) communication systems has spurred interest in exploiting large antenna apertures and higher frequency bands to meet increasing data throughput and spectral efficiency demands \cite{liu2024near}. As the aperture size increases and the wavelength decreases, the near-field region expands and the spherical-wave nature of electromagnetic propagation becomes significant. Consequently, future wireless systems are more likely to operate in these near-field regimes \cite{zhang2020prospective}, which has increased research interest in near-field communications.

Unlike far-field scenarios, where beamforming is mainly used to steer energy along specific angular directions, the near-field enables \emph{beam focusing} that leverages both angular and distance domains. This capability permits spatial multiplexing of users that share the same angular direction if they are located at different distances. Such an approach could be particularly useful in ultra-dense scenarios where many users are aligned in the same direction, a situation that challenges traditional far-field beamforming methods. Several studies have investigated spatial multiplexing based on the distance domain; for instance, \cite{zhang2022beam} and \cite{li2023near} demonstrated through simulations that two collinear users can be spatially multiplexed in a LoS near-field channel. However, these works did not delve into a theoretical explanation of channel characteristics for the effectiveness of such multiplexing.
\textcolor{black}{To address this gap, \cite{wu2023multiple} and \cite{wu2023enabling} investigated the channel correlation between collinear users in the near-field region.
Once the correlation falls below a certain threshold, spatial multiplexing becomes feasible.}
Based on the closed‐form channel‐correlation model in~\cite{wu2023multiple}, Kosasih \emph{et al}.~\cite{kosasih2024finite} introduced the concept of a finite beam depth for spatial multiplexing in the distance domain. In the conventional far‐field regime, the beam depth tends to diverge to infinity—resulting in very high correlation between collinear users and hence precluding distance‐domain multiplexing. By contrast, the near‐field region can yield a finite beam depth, which is equivalent to low spatial correlation between two collinear users, thereby enabling spatial multiplexing. 
\textcolor{black}{This finite beam depth has been characterized analytically for a variety of aperture geometries, including rectangular and circular apertures ~\cite{kosasih2024finite}, modular apertures~\cite{kosasih2024achieving,li2024multi}, and sparse arrays~\cite{chen2024near,zhou2024sparse}. These studies highlight the strong potential for distance-domain spatial multiplexing across diverse antenna architectures.}

In this work, we investigate the distance-domain DoF, which can provide many useful insights into the spatial multiplexing performance and channel estimation. For point-to-point (P2P) MIMO, the distance-domain DoF establish the maximum number of parallel non-interfering channels between Tx array and Rx array, where Rx array spans solely over the distance domain. In the context of multi-user MIMO (MU-MIMO), the distance-domain DoF can establish the maximum number of users that can be spatially multiplexed with minimal interference when they are at the same angle but different distance. In the context of channel estimation, the distance-domain DoF can represent the minimum number of samples/sweeping beams to recover the distance-dependent channel, assuming known user's angle.
As a result, establishing this limit sharpens our understanding of distance-domain beamforming and channel estimation.
While recent studies~\cite{zhang2022beam,li2023near,wu2023multiple,wu2023enabling,kosasih2024finite} demonstrate the feasibility and benefits of distance-domain spatial multiplexing, they do not characterize the {theoretical} limit on the spatial DoF available along the distance dimension. 

To obtain \textcolor{black}{a} closed-form expression of the distance-domain DoF, we consider an equivalent point-to-point MIMO (P2P-MIMO) channel between a base station as a continuous-aperture transmit array of arbitrary shape and a continuous-aperture arbitrarily long receive array that spans solely over the distance domain. 
The continuous array can be arbitrarily long, capturing scenarios where users span a wide range of distances. Moreover, we allow the base-station aperture to have arbitrary shape, reflecting practical design flexibility.
{We focus on the LoS component, which typically dominates in mmWave/THz bands due to significant NLoS attenuation \cite{priebe2013stochastic}. Since multipath scattering typically increases channel rank by introducing virtual sources \cite{tse2005fundamentals}, our LoS analysis establishes a fundamental DoF lower bound characterizing performance in sparse-scattering environments.}

\textcolor{black}{\subsection{Related Work}}

The DoF of near-field channels have been extensively investigated in point-to-point (P2P) MIMO settings; therefore, it is important to delineate how our problem relates to and differs from that literature. In general, most analyses consider geometries in which the receive (Rx) aperture spans primarily the \emph{angular} domain of the transmit (Tx) aperture. 
Prior work has largely focused on configurations where one aperture covers the angular extent of the other, including linear arrays~\cite{xie2023performance,pu2014effects,torkildson2011indoor,kanatas2025deterministic}, rectangular arrays~\cite{hu2018beyond,dardari2020communicating,sun2021small,pizzo2022landau,do2023parabolic,gong2024holographic,wang2025analytical,guo2025impact,hamza2025spatial,hussain2026analyzing}, circular array~\cite{xu2017degrees} and more general apertures~\cite{yuan2021electromagnetic,shen2023electromagnetic,xu2023exploiting,miller2000communicating,pizzo2022nyquist,gustafsson2025degrees}.
\textcolor{black}{Mathematically, spatial DoF evaluation in near-field configurations heavily relies on integral operator kernels. The literature primarily utilizes two kernel-based approaches:
\begin{enumerate}
    \item Estimating the effective DoF via the trace of the spatial correlation kernel—typically an energy ratio between its squared trace and Frobenius norm \cite{xie2023performance, wang2025analytical, yuan2021electromagnetic, shen2023electromagnetic, gustafsson2025degrees, hamza2025spatial, hussain2026analyzing}.
    \item Exploiting the band-limited characteristics of the communication kernel via Landau's eigenvalue theorem \cite{pizzo2022landau, do2023parabolic} or Prolate Spheroidal Wave Functions (PSWFs) \cite{pu2014effects, torkildson2011indoor, guo2025impact}.
\end{enumerate}
While these studies successfully apply kernel methods for standard user distributions, isolating the pure distance-domain DoF for collinear users requires a fundamentally different kernel transformation, which is the focus of this work.}

Compared to this extensive treatment of angular-domain DoF, the \emph{distance}-domain DoF has received comparatively little attention. Some works do consider cases where the Rx aperture varies mainly along the distance domain of the Tx array \cite{decarli2021communication,ruiz2023degrees,ding2022degrees,chen2025degrees}. In~\cite{decarli2021communication,ruiz2023degrees}, the DoF between two rectangular apertures with mutually perpendicular orientations is analyzed. Although the Rx aperture in these studies spans the distance domain, it also extends over the angular domain. Hence, the DoF expression in \cite{decarli2021communication,ruiz2023degrees} is a combination of the distance-domain and angular-domain DoF, which prevents us from understanding the true values of the distance-domain DoF. In~\cite[Sec.~IV-B]{ding2022degrees}, two linear arrays with perpendicular orientations are considered; this setup can capture distance-domain DoF for collinear users when the base station is modeled as a linear array, but it does not generalize to arbitrary Tx aperture shapes. Moreover, \cite{chen2025degrees} derives a closed-form DoF expression for a rectangular Tx aperture and two collinear users at fixed positions. While insightful, the restriction to two predetermined users does not address the broader question studied here: determining the distance-domain DoF, equivalently maximum number of collinear users that can be spatially multiplexed solely via distance separation. Finally, all the works in \cite{decarli2021communication,ding2022degrees,ruiz2023degrees,chen2025degrees} assume Tx–Rx separations larger than the arrays dimensions, which does not reflect scenarios with collinear users distributed over a wide \emph{distance} span.

\subsection{Contributions}

We investigate and derive a closed-form interpretable expression of the \emph{distance-domain} DoF, which can provides some insights into the multiplexing gain performance and channel estimation.
Our main contributions are as follows.

\subsubsection{System Model}
We adopt a geometric framework that can capture the distance-domain DoF:
\begin{itemize}
    \item \textbf{Source (represents base station):} The source (Tx array) may have \emph{arbitrary shape} and may include \emph{gaps} (capturing modular and sparse arrays). The only constraint is its inner and outer size, specified by the maximum radial extent from the array-center coordinate to its boundary.
    \item \textbf{Observation region (represents users):} The observation region (Rx array) is a collinear array whose length can be arbitrarily large, which can represent multiple collinear users separated over a long distance span.
    \item \textbf{Continuous model:} We idealize both Tx and Rx as continuous-aperture (CAP) arrays. While being hardware-challenging, it serves as an information-theoretic benchmark: the CAP DoF establishes an \emph{upper bound} on the DoF of any sampled (discrete) implementation.\footnote{Extension of this work, i.e., the problem of sampling the continuous source and observation to maximize the distance-domain DoF, is under preparation for publication \cite{duong2025sampling}.} This abstraction also yields clean insight into multi-user MIMO and channel-estimation limits.
\end{itemize}
In contrast, the literature focuses on P2P-MIMO with two specific, finite arrays (e.g., ULA/URA/UCA) and Rx size is smaller than the Tx–Rx separation, which cannot model the distance-domain DoF with wide distance span.

\subsubsection{Technical Contributions}
For clarity and to expose geometry-driven effects step by step, we analyze three canonical scenarios:
(i) a single-piece Tx aperture with a \emph{broadside} Rx array;
(ii) a \emph{non-broadside} Rx array; and
(iii) a \emph{modular} Tx aperture.
The key results are:
\begin{itemize}
    \item \textbf{Single-piece Tx aperture \& broadside Rx array:} Using a scalar field model with a uniform-wavefront approximation, we cast the channel as a linear integral operator. Since the native kernel is non-convolutional, we apply a change of variables to obtain an \emph{equivalent} operator with a Hermitian convolution kernel, enabling eigenvalue analysis via Landau’s theorem. This yields a closed-form DoF:
    \begin{align}
        \mathrm{DoF}
        \;=\;
        \frac{\big(p_{\max}^{2}-p_{\min}^{2}\big)\,\big(r_{\min}^{-1}-r_{\max}^{-1}\big)}{2\lambda}
        \;+\; {O}(1),
        \label{eqn:proposed_distance_domain_DoF}
    \end{align}
    where $p_{\min}$ and $p_{\max}$ denote the minimum and maximum \emph{Tx} radial distances from the array center, and $r_{\min}$ and $r_{\max}$ denote the minimum and maximum \emph{Rx} distances (along the collinear array) from the same center. The expression shows that the DoF depends only on the \emph{extreme edges} of the apertures, not on the detailed Tx shape and aperture's area.
    \item \textbf{Non-broadside Rx array:} For a general pointing, we orthogonally project the Tx aperture onto the plane perpendicular to the Rx array direction, producing a \emph{projected} Tx aperture. This reduces the non-broadside geometry (Tx, Rx) to an equivalent broadside geometry (projected Tx, Rx), preserving the DoF and yielding a simple geometric rule for evaluation.
    \item \textbf{Modular Tx array:} We extend the framework to modular apertures. We show that introducing a central gap does \emph{not} substantially degrade the distance-domain DoF. Moreover, for a fixed total Tx area, a suitably designed modular aperture can achieve a \emph{higher} distance-domain DoF than a single-piece aperture.
\end{itemize}

\subsubsection{Distinction from Angular-Domain DoF Results}
\label{sec:novelty_distinction}

It is worth to distinguish the \emph{distance-domain} DoF derived in \eqref{eqn:proposed_distance_domain_DoF} from the well-established \emph{angular-domain} DoF results found in the literature (e.g., Pizzo \emph{et al.} \cite{pizzo2022landau} and Ruiz-Sicilia \emph{et al.} \cite{ruiz2023degrees}). Given the extensive literature on angular-domain limits, one might intuitively assume the our proposed distance-domain DoF is merely a special case of the angular-domain DoF, obtainable by adjusting orientation parameters in the standard formula \cite{torkildson2011indoor,pizzo2022landau,miller2000communicating,xie2023performance,pu2014effects,ruiz2023degrees,do2023parabolic,xu2017degrees,yuan2021electromagnetic,xu2023exploiting,guo2025impact}:
\begin{align}
    \mathrm{DoF}_{\text{angular}} \approx \frac{\alpha A_t A_r}{(\lambda D)^n}.
    \label{eqn:angular_DoF_misconception}
\end{align}
where \(\alpha\) is coefficient determined via orientation of Tx and Rx arrays, \(A_t\) and \(A_r\) are the physical areas of Tx and Rx arrays, \(D\) is the Tx–Rx separation (treated as a single representative distance), \(n\) depends on aperture dimensionality.

It can be demonstrated that \eqref{eqn:angular_DoF_misconception} from \cite{pizzo2022landau, ruiz2023degrees} cannot yield the distance-domain DoF for two fundamental reasons:
\begin{enumerate}
    \item \textbf{Dependence on Area vs. Radial Extent:} The angular-domain DoF in \eqref{eqn:angular_DoF_misconception} scales linearly with the aperture areas ($A_t, A_r$). In stark contrast, our derived distance-domain DoF \eqref{eqn:proposed_distance_domain_DoF} depends on the \emph{difference of squares} of the radial extents ($p_{\max}^2 - p_{\min}^2$) and difference of inverse distance ($r^{-1}_{\min} - r^{-1}_{\max}$) and not on the aperture areas.
    \item \textbf{Distance Spanning:} Angular-domain models typically assume the Rx array dimensions are small relative to the transmission distance, effectively treating the Tx-Rx separation $D$ as a constant. Our model captures the \emph{wide-span} regime where the Rx array extends from $r_{\min}$ to $r_{\max}$. Attempting to apply \eqref{eqn:angular_DoF_misconception} by letting the Rx area grow along the distance axis leads to unbounded DoF growth, whereas our result correctly approaches a finite saturation limit as $r_{\max} \to \infty$.
\end{enumerate}

The remainder of the paper is organized as follows. In Section II, we describe the system model, including the geometrical setup of Tx and Rx arrays, the near-field channel model. Section III presents the definition of distance-domain DoF and its implications in practice. Section IV presents the mathematical framework and derives a closed-form expression for the distance-domain DoF in the ideal case of Tx array and Rx array, with Rx array located in the broadside of Tx array \footnote{Parts of this paper including Section II, III-A and IV was presented in IEEE Globecom 2025 \cite{duong2025spatial}.}. 
Section V extends the analysis to the non-broadside configuration of the Rx array. Section VI extends the analysis to the modular Tx array.
Finally, Section VII concludes the paper.

%% file: sec2_1_channelModel_v4.tex
\section{System Model}

\subsection{Geometrical Setup}

\textcolor{black}{To rigorously isolate the distance-domain {DoF} from the angular-domain DoF, our {LoS} {MIMO} channel assumes all receive (Rx) elements align along a single angle, differing only in radial distance. This mathematically reduces the observation region to a {1D} linear array, because a {2D} Rx region would inherently entangle the distance- and angular-domain {DoF}. Conversely, we assume an arbitrary {2D} shape for the transmit (Tx) array to maximize the generality of our analytical framework across diverse base station architectures.} The detailed system parameters are defined as follows.

\begin{figure}[ht]
    \centering
    \includegraphics[width=5cm]{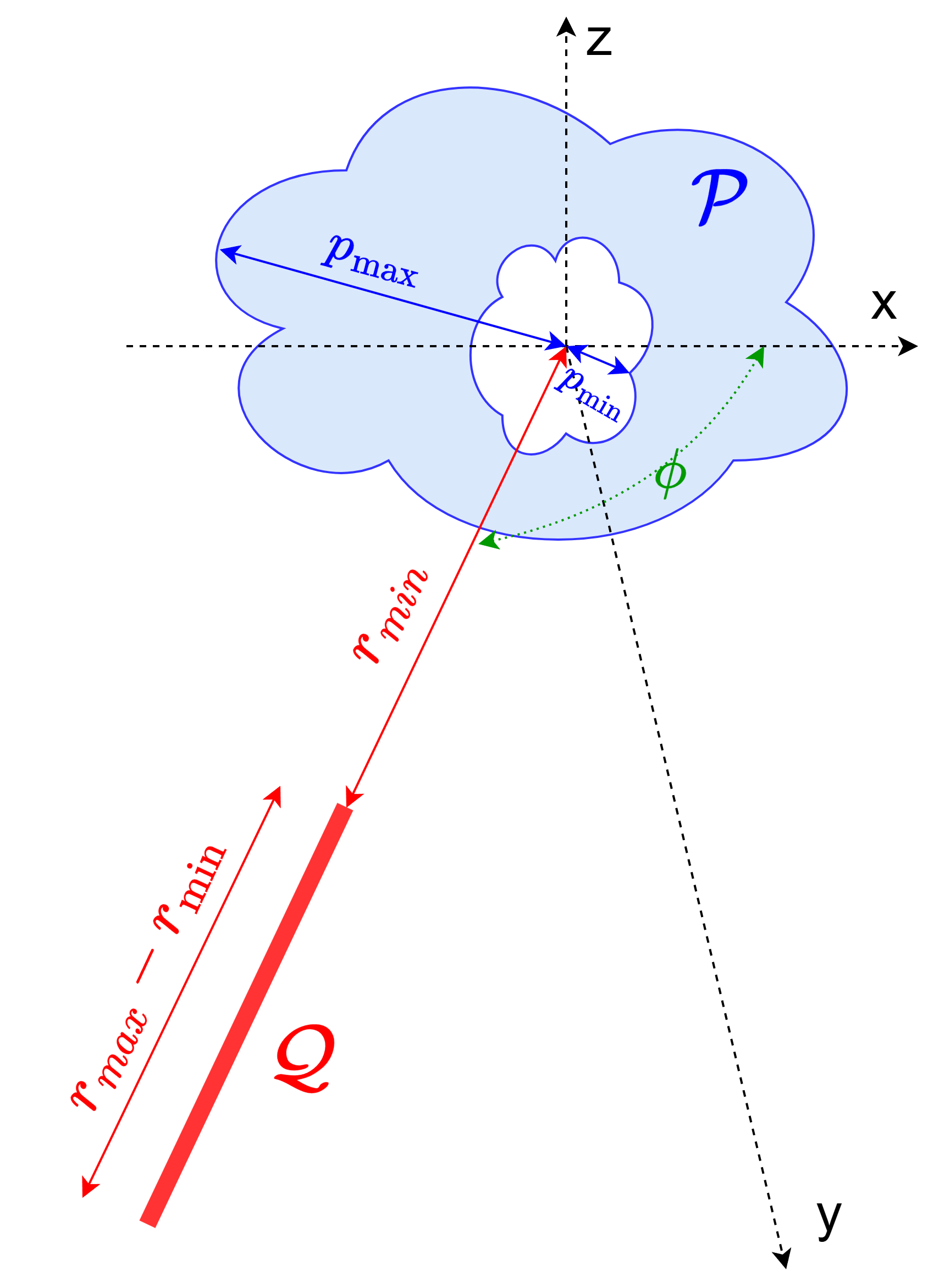}
    \caption{Geometrical setup of the Tx array \(\mathcal{P}\) (with an example of a gap inside \( \mathcal{P} \)) and the Rx broadside linear array \(\mathcal{Q}\).}
    \label{fig:broadside_collinear_elements_2}
\end{figure}

\subsubsection{Tx array (Base station)}
Let \( \mathcal{P} \) denote the Tx-aperture region and \( \mathbf{p}\in\mathcal{P} \) a point on it. For simplicity, we assume \( \mathcal{P} \) lies on the \(xz\)-plane, so any point has the form \( \mathbf{p}=[p_1,\,0,\,p_3]^{\mathsf T} \). 
Assumptions on \( \mathcal{P} \) are:
\begin{itemize}
    \item \textbf{General shape:} \( \mathcal{P} \) is an arbitrary (at most two-dimensional) aperture and may include internal gaps (e.g., modular or sparse configurations), cf. Fig.~\ref{fig:broadside_collinear_elements_2}.
    \item \textbf{Finite boundary}: We assume \( \mathcal{P} \)'s size to be finite, bounded by \( p_{\min} \leq \| \mathbf{p} \| \leq p_{\max} \). In the words, \( p_{\min} \) and \( p_{\max} \) are the distances from its extreme inner and outer edges to \( \mathbf{O} \):
    \begin{align}
        p_{\min} \triangleq \min_{\mathbf{p}\in\mathcal{P}} \|\mathbf{p}\|, \qquad
        p_{\max} \triangleq \max_{\mathbf{p}\in\mathcal{P}} \|\mathbf{p}\|.
        \label{eq:extremes}
    \end{align}
    \item \textbf{Element spacing:} The element spacing is infinitesimal (or sufficiently small) so that an areal integral over \( \mathcal{P} \) accurately models the channel and DoF. Hence, Tx array is considered to have continuous aperture.
    \item \textbf{Connectedness:}
    \begin{itemize}
        \item \textbf{Single piece:} \( \mathcal{P} \) is connected, and there exists a line segment within \( \mathcal{P} \) joining the extreme points achieving \( p_{\min} \) and \( p_{\max} \). This case is analyzed in Section~\ref{sec:framework_spatial_DoF}.
        \item \textbf{Modular aperture:} \( \mathcal{P} \) comprises multiple disjoint sub-apertures. This extension is treated in Section~\ref{sec:modular_aperture}.
    \end{itemize}
    \item \textbf{Off-center possibility:} \( \mathbf{O} \) need not coincide with the geometric center of \( \mathcal{P} \) (and may even lie outside \( \mathcal{P} \)); in such cases \( p_{\min}\neq 0 \), cf. Fig.~\ref{fig:broadside_collinear_elements_2}.
\end{itemize}

\subsubsection{Rx linear array (can represent collinear users)}
Let \( \mathcal{Q} \) denote the Rx-array axis, and parameterize any point \( \mathbf{q}\in\mathcal{Q} \) as
\begin{align}
    \mathbf{q} = r\,\mathbf{k}, \label{eq:q_position}
\end{align}
where \( r=\|\mathbf{q}\| \) and the unit direction vector
\begin{align}
    \mathbf{k} = \big[\cos\phi\,\cos\theta,\; \sin\phi\,\cos\theta,\; \sin\theta\big]^{\mathsf T}.
\end{align}
For simplicity, we assume the Rx array lies on the xy-plane, so \( \theta = 0 \). In the broadside case, \( \mathbf{k}=[0,1,0]^{\mathsf T} \), so \( \mathcal{Q} \) coincides with the \(y\)-axis (see Fig.~\ref{fig:broadside_collinear_elements_2}).

Assumptions on the Rx array \( \mathcal{Q} \) are:
\begin{itemize}
    \item \textbf{Element spacing:} The element spacing is infinitesimal (or sufficiently small) so that an integral operator can accurate capture DoF. Hence, Rx array is considered to be CAP array.
    \item \textbf{Fixed direction}: \( \mathbf{k} \) is fixed for all \( \mathbf{q}\in\mathcal{Q} \), so only distance-domain DoF is involved and not angular-domain DoF.
    \item \textbf{Infinite boundary}: We assume Rx array is bounded by:
    \begin{align}
        r \in [\,r_{\min},\, r_{\max}\,], \label{eq:bounded_range}
    \end{align}
    where the \emph{distance span} \( r_{\max}-r_{\min} \) may be arbitrarily large and $r_{\max}-r_{\min}$ can possibly approach infinity. This contrasts with common assumptions under which \( r_{\max}-r_{\min} \ll r_{\max} \), so the Tx–Rx separation is effectively a single distance~\cite{do2023parabolic,pizzo2022landau}. The infinite boundary is require to yield an upper bound on the distance-domain DoF for multi-user MIMO or channel estimation, where the distance span can be large. Moreover, $r_{\max}$ may approach the far field. Since the Fresnel approximation improves with increasing distance, our framework naturally covers joint near-field and far-field regimes.
\end{itemize}

\subsection{Near-Field Channel Model}

We consider a LoS channel between the Tx CAP array \( \mathcal{P} \) and the Rx CAP linear array \( \mathcal{Q} \).
A full vector-wave description employs the free-space {dyadic} Green’s function \cite{wang2025analytical}:
\begin{equation}
\mathbf{E}(\mathbf{q})
= j\omega\mu \int_{\mathcal{P}}
\Big(\overline{\overline{\mathbf I}} + \frac{1}{k^{2}}\nabla\nabla\Big) \, h(\mathbf{q},\mathbf{p}) \cdot \mathbf{J}(\mathbf{p}) \, \mathrm d\mathbf{p},
\label{eq:dyadic}
\end{equation}
where \(k=2\pi/\lambda\), \(h(\mathbf{q},\mathbf{p})\) is the scalar free-space Green’s function \cite{wang2025analytical}
\begin{align}
h(\mathbf{q},\mathbf{p}) \;=\; \frac{e^{-jk\lVert \mathbf{q}-\mathbf{p}\rVert}}{4\pi\,\lVert \mathbf{q}-\mathbf{p}\rVert},
\end{align}
\(\mathbf{J}(\mathbf{p})\) is the transmit surface current density, and the differential operator acts with respect to \(\mathbf{q}\). 
The dyadic term accounts for polarization, but retaining it complicates the analysis and obscures closed-form DoF expressions. To obtain a closed-form and interpretable DoF expression, we simply adopt the scalar kernel \(h(\mathbf q,\mathbf p)\), which is accurate under standard array conditions: (i) single fixed Tx/Rx polarization; (ii) each Rx point lies in the radiative field of each differential Tx element—i.e., the Tx–Rx spacing exceeds the element Fraunhofer distance \(r \gg 2a^{2}/\lambda\) for element size \(a\); and (iii) polarization varies slowly across \(\mathcal{P}\) for a fixed look direction.

\paragraph*{Uniform-amplitude and Fresnel approximations}
When the minimum Tx–Rx distance is larger than the Tx-aperture size (often termed the \emph{Bjohnson distance}~\cite{kosasih2024finite}), each differential Tx element illuminates the Rx line with nearly uniform amplitude. Under this \emph{uniform-amplitude} assumption and the Fresnel approximation, for \(\mathbf{q}=r\,\mathbf{k}\) with \(\|\mathbf{k}\|=1\),
\begin{align}
\lVert \mathbf{q}-\mathbf{p}\rVert
&
\;\approx\;
r \;-\; \mathbf{k}^{\mathsf  T}\mathbf{p}
\;+\; \frac{\lVert\mathbf{p}\rVert^{2}-\big(\mathbf{k}^{\mathsf T}\mathbf{p}\big)^{2}}{2r}, \label{eq:fresnel_distance}
\\
\frac{1}{\lVert \mathbf{q}-\mathbf{p}\rVert}
&\approx \frac{1}{r}, \nonumber
\end{align}
which yields the scalar kernel
\begin{align}
h(\mathbf{p},\mathbf{q})
\;\approx\;
\frac{1}{r}\,
\exp\!\left(
-\frac{j2\pi}{\lambda}
\Big[
r \;-\; \mathbf{k}^{\mathsf T}\mathbf{p}
\;+\; \frac{\lVert\mathbf{p}\rVert^{2}-\big(\mathbf{k}^{\mathsf T}\mathbf{p}\big)^{2}}{2r}
\Big]
\right).
\label{eq:channel_coef_norm}
\end{align}
The global phase factor \(e^{-j2\pi r/\lambda}\) is common to all \(\mathbf{p}\in\mathcal{P}\) and can be dropped without affecting the DoF.

%% file: sec2_2_DoF_definition_v4.tex
\section{Distance-Domain Degrees of Freedom and Implications}

\subsection{Distance-domain DoF in Point-to-point MIMO Channel}
For spatially discrete arrays, the LoS DoF is often taken as the number of dominant eigenvalues of
\(\mathbf V=\mathbf H^{\mathsf H}\mathbf H\), where \(\mathbf H\) is the discrete channel matrix.
For CAP arrays with infinitesimal spacing, the Rx-side correlation is the self-adjoint, positive, compact integral operator
\begin{align}
    (\mathcal V \Phi)(\mathbf q)
    = \int_{\mathbf q' \in \mathcal Q} v(\mathbf q,\mathbf q')\, \Phi(\mathbf q')\, \mathrm d\mathbf q',
    \label{eq:operator_V}
\end{align}
with Hilbert–Schmidt kernel
\begin{align}
    v(\mathbf q,\mathbf q')
    = \int_{\mathbf p \in \mathcal P} h^{*}(\mathbf p,\mathbf q)\, h(\mathbf p,\mathbf q')\, \mathrm d\mathbf p.
    \label{eq:kernel_def}
\end{align}
By Mercer’s theorem, there exist orthonormal eigenfunctions \( \{e_n\}_{n\ge1}\subset L^2(\mathcal Q)\) and non-increasing eigenvalues \( \{\lambda_n\}_{n\ge1}\) such that
\begin{align}
    v(\mathbf q,\mathbf q') = \sum_{n=1}^{\infty} \lambda_n\, e_n(\mathbf q)\, e_n^{*}(\mathbf q'), \\
    \lambda_n e_n(\mathbf q) = \int_{\mathcal Q} v(\mathbf q,\mathbf q')\, e_n(\mathbf q')\, \mathrm d\mathbf q'.
    \label{eq:mercer}
\end{align}
Theoretically, the DoF can be infinite since the eigenvalues can decay exponentially, but remains strictly greater than zero. Practically, \emph{effective} DoF is defined via a threshold \(\varepsilon>0\) as
\begin{align}
    N_{\varepsilon} \triangleq \#\{\, n : \lambda_n \ge \varepsilon \,\}.
    \label{eq:dof_def}
\end{align}
\textcolor{black}{In practical systems, the choice of $\epsilon$ is determined by the environmental noise floor, ensuring that the spatial sub-channels associated with the counted eigenvalues have sufficient strength for reliable communication. Additionally, the effective {DoF} calculation is notably stable with respect to the choice of $\epsilon$. Because the near-field eigenvalues experience a exponential decay following the cutoff index, adjusting the threshold from $\epsilon_1$ to $\epsilon_2$ results in merely a marginal logarithmic shift, scaling as $\mathcal{O}(\log(\epsilon_2/\epsilon_1))$.}

\subsection{Implications of Distance-Domain DoF for MU-MIMO and Channel Estimation}
\begin{figure}[ht]
    \centering
    \begin{subfigure}[b]{0.47\textwidth}
        \centering
        \includegraphics[width=8.5cm]{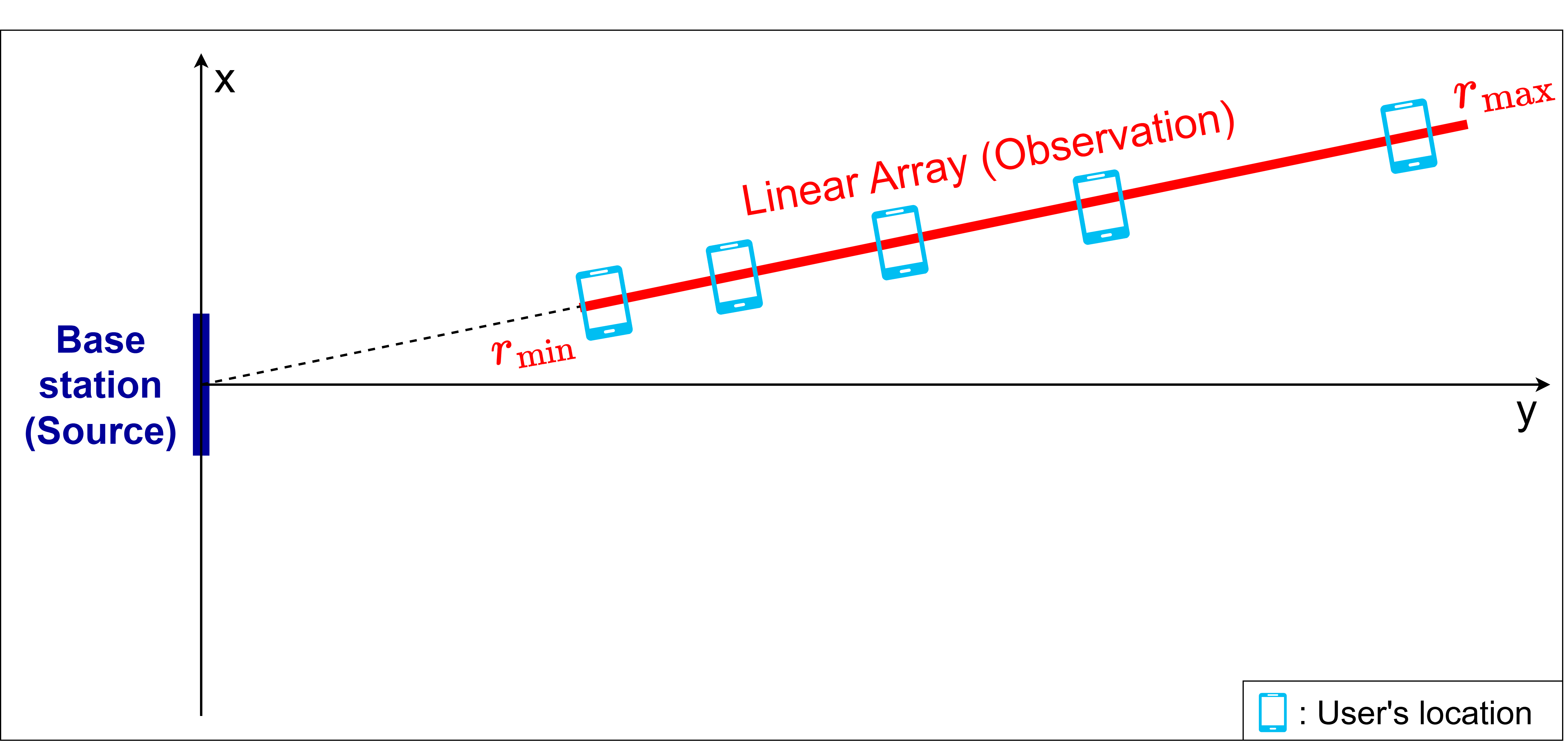}
        \caption{\textbf{Maximum number of users that can be spatially multiplexed via distance separation.}}
        \label{fig:polar_domain_v3}
    \end{subfigure}
    \vfill 
    \begin{subfigure}[b]{0.47\textwidth}
        \includegraphics[width=8.5cm]{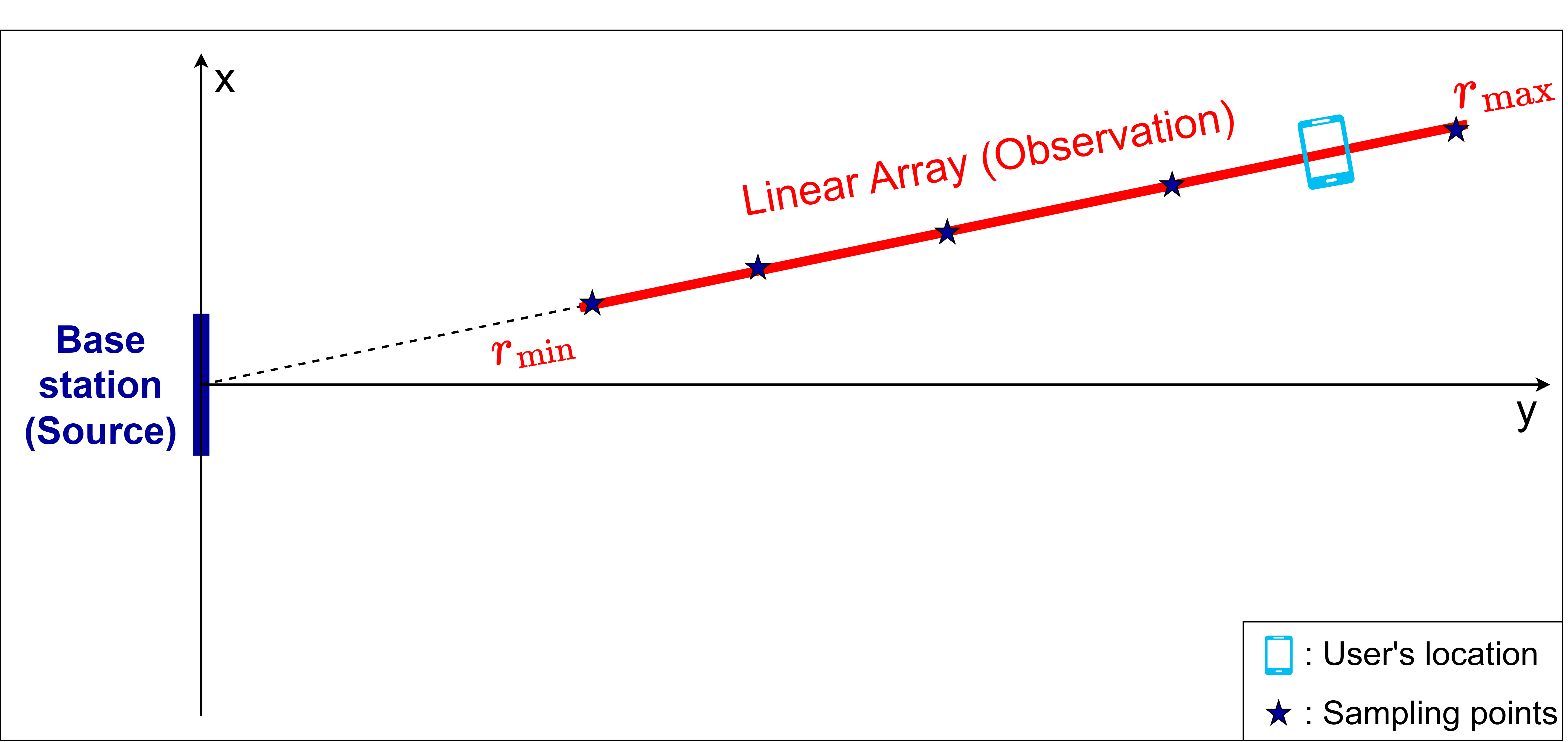}
        \caption{\textbf{Minimum number of distance-domain samples to recover the user's channel information, given known user's angle information.}}
        \label{fig:polar_domain_v3b}
    \end{subfigure}
    \caption{Implication of distance-domain DoF for users distributed within the distance span $[r_{\min}, r_{\max}]$. The linear array (red) represents the region where single-antenna users can be located.}
    \label{fig:distance_DoF_implications}
\end{figure}

Clearly, the DoF of the Tx array and collinear Rx array is not necessarily useful in the context of P2P-MIMO, since it is not practical to constrain the Rx array to lie on a fixed line through the BS. However, the DoF of this P2P-MIMO channel can provide some useful insights into the context of multi-user MIMO:

\subsubsection{Upper bound on MU–MIMO DoF (same angle, distance-separated)}
Let \(K\) collinear users lie at the the same angle but different distances bounded by $[r_{\min}, r_{\max}]$, as shown in Fig.~\ref{fig:polar_domain_v3}. 
Truncating \eqref{eq:mercer} at \(N\) terms (e.g., \(N=N_{\varepsilon}\)) gives:
\begin{align}
    v(\mathbf q,\mathbf q') \approx \sum_{n=1}^{N} \lambda_n e_n(\mathbf q)e_n^{*}(\mathbf q').
\end{align}
By denoting \(\{\mathbf q_k\}_{k=1}^{K}\subset\mathcal Q\) as the positions of the \(K\) users, the user's channel correlation matrix \( \mathbf{V} \) is a discrete Gram matrix, i.e., \(\mathbf V=[v(\mathbf q_i,\mathbf q_j)]_{i,j=1}^{K}\), and admits the truncated spectral (low-rank) factorization
\begin{align}
    \mathbf{V} & \approx \mathbf{E} \mathbf{\Lambda} \mathbf{E}^{\text{H}}, \\
    \mathbf{E} & = [\, \mathbf{e}_1\ \cdots\ \mathbf{e}_N \,], \ \mathbf{e}_n = [\, e_n(\mathbf{q}_1), \dots, e_n(\mathbf{q}_K) \,]^{\text{T}},\\
    \mathbf{\Lambda} & = \text{diag}(\lambda_1, \dots, \lambda_N).
    \label{eq:discrete_approx}
\end{align}
Hence
\begin{align}
    \text{rank}(\mathbf{E} \mathbf{\Lambda} \mathbf{E}^{\text{H}}) \le \min\{\text{rank}(\mathbf{E}), \text{rank}(\mathbf{\Lambda})\} \le N \le N_{\varepsilon}.
\end{align}
Therefore, the CAP DoF upper-bounds the DoF of any discrete MU–MIMO realization sampled along that angle. If there are more users than the available distance-domain DoF between $[r_{\min}, r_{\max}]$ \((K>N_{\varepsilon})\), the channel matrix \(\mathbf H\) necessarily has small singular values, leading to ill-conditioning and degraded spatial multiplexing gain of MMSE/ZF-type precoders whose performance depends on the singular-value distribution. 
\textcolor{black}{Consequently, this rank deficiency imposes a strict physical ceiling on the achievable multi-user sum-capacity; because inter-user interference cannot be effectively mitigated in the spatial domain when $K > N_{\varepsilon}$, the system capacity will inevitably saturate or degrade.}

\subsubsection{Minimum distance-domain sampling for near-field training}

When the user angle is known but the user's unknown distance lies in \( [r_{\min},r_{\max}] \) as shown in Fig.~\ref{fig:polar_domain_v3b}, one fundamental question arises, i.e., the minimum number of distance-domain samples required to recover the channel. To answer this, we can analyze the DoF of the MIMO channel between the Tx array and the Rx linear array that spans over \( [r_{\min},r_{\max}] \).

Specifically, given the orthonormal eigenfunctions \( \{e_n\}_{n\ge1} \) and the nonincreasing eigenvalues \( \{\lambda_n\}_{n\ge1} \) of \eqref{eq:operator_V}, the scalar Green kernel admits the truncated Hilbert–Schmidt expansion
\begin{align}
    h(\mathbf p,\mathbf q)
    \;\approx\;
    \sum_{n=1}^{N_{\varepsilon}} \sqrt{\lambda_n}\, f_n(\mathbf p)\, e_n^{*}(\mathbf q),
\end{align}
where \( \mathbf q \) is the (unknown) user location and \( \mathbf p\in\mathcal P \) is a Tx-aperture point. For a spatially discrete Tx array sampled at \( \{\mathbf p_m\}_{m=1}^{M} \), the channel vector \( \mathbf h(\mathbf q) \in \mathbb{C}^{M \times 1} \) at user's location \( \mathbf{q} \) given by:
\begin{align}
    \mathbf h(\mathbf q)
    \;=\;
    \big[\, h(\mathbf p_1,\mathbf q),\ldots,h(\mathbf p_M,\mathbf q) \,\big]^{\mathsf T}
\end{align}
admits the finite-basis approximation
\begin{align}
    \mathbf h(\mathbf q)
    & \;\approx\;
    \sum_{n=1}^{N_{\varepsilon}}
    \big(\sqrt{\lambda_n}\, e_n^{*}(\mathbf q)\big)\,\mathbf f_n,
    \\
    \mathbf f_n & \;=\; \big[\, f_n(\mathbf p_1),\ldots,f_n(\mathbf p_M) \,\big]^{\mathsf T}.
\end{align}
With sufficiently dense uniform sampling of \( \{\mathbf p_m\}_{m=1}^{M} \), the set \( \{\mathbf f_n\}_{n=1}^{N_{\varepsilon}} \) is (approximately) orthonormal (since the set of continuous basis\( \{ f_n(\mathbf p) \}_{n=1}^\infty \) is orthonormal).

As a result, \( \mathbf h(\mathbf q) \) lies, up to the accuracy set by \( \varepsilon \), in an \( N_{\varepsilon} \)-dimensional subspace. Consequently, any channel estimation technique that aims to approximate \( \mathbf h(\mathbf q) \) with a \( C \)-dimensional subspace by using the following codebook \( \mathbf W \in \mathbb C^{M\times C} \), i.e., \cite{cui2022channel}
\begin{align}
    \hat{\mathbf{h}}(\mathbf q) \approx \mathbf W \, \hat{\mathbf{d}}(\mathbf q),
\end{align}
must use 
\[
C \;\ge\; N_{\varepsilon}
\]
samples/beams to obtain distance-domain information in the interval \( [r_{\min}, r_{\max}] \), assuming that the angle's information is already known.

%% file: sec3_1_framework_v4.tex
\section{Proposed Distance-Domain DoF Derivation with Broadside Case Receive Array and Single-Piece Transmit Array}
\label{sec:framework_spatial_DoF}

\subsection{Formulation of Integral Operator with Convolution Kernel}

Since the Rx array \( \mathcal{Q} \) is one-dimensional, every point \( \mathbf{q} \in \mathcal{Q} \) can be uniquely represented by its distance \( r \) from the center coordinate. We can simplify the operator \( (\mathcal{V} \Phi)(\mathbf{q}) \) as follows:
\begin{align}
	( {\mathcal{V}} \Phi)(r) = \int_{r_{\min}}^{r_{\max}} {v}(r,r')\, \Phi(r')\, dr', \ \forall r \, \in [r_{\min}, r_{\max}],
	\label{eq:operator_V_simplified}
\end{align}
where substituting (\ref{eq:channel_coef_norm}) into (\ref{eq:kernel_def}) results in the kernel \( {v}(r,r') \) as follows:
\begin{align}
	& {v}(r,r') \approx \nonumber \\
    & \frac{1}{rr'}\iint_{\mathcal{P}} \exp\!\Bigg(-\frac{j2\pi}{\lambda}\frac{\|\mathbf{p}\|^2 - (\mathbf{k}^T \mathbf{p})^2}{2}\Big(\frac{1}{r}-\frac{1}{r'}\Big)\Bigg)d\mathbf{p},
	\label{eq:kernel_after_approx}
\end{align}
which only retains the quadratic distance-dependent phase since the linear part is eliminated.

In the case of broadside Tx array where \(\mathbf{k}^T \mathbf{p} = 0, \  \forall \, \mathbf{p} \in \mathcal{P}\), we can simplify the kernel \(\check{v}(r,r')\) as follows:
\begin{align}
	{v}(r,r') &\approx \frac{1}{rr'}\iint_{\mathcal{P}} \exp\!\Bigg(-\frac{j2\pi}{\lambda}\frac{\|\mathbf{p}\|^2}{2}\Big(\frac{1}{r}-\frac{1}{r'}\Big)\Bigg)d\mathbf{p}.
	\label{eq:kernel_broadside}
\end{align}

\textcolor{black}{Note that its kernel \( {v}(r,r') \) in (\ref{eq:kernel_after_approx}) is not Hermitian kernel, since the amplitude \( {1}/{rr'} \) is not constant and \( r \) can vary over a large distance, and the phase difference is a function of $1/r - 1/r'$ rather than the shift-invariant difference $1/r - 1/r'$. This is different from other P2P MIMO channel, where the element-wise distance between Tx and Rx arrays are relatively constant. To facilitate the DoF analysis via Landau's Theorem \cite{landau1975szego}, we perform the following variable change:}
\begin{align}
	 \zeta = \frac{1}{r} \quad \text{and} \quad \zeta' = \frac{1}{r'}.
	\label{eq:inv_distance}
\end{align}
Given new variable, we define the operator \( \mathcal{G} \):
\begin{align}
	(\mathcal{G} \Phi)( \zeta) = \int_{r^{-1}_{\max}}^{r^{-1}_{\min}} g(\zeta,\zeta')\, \Phi( \zeta')\, dt', \quad \zeta \in \Big[r^{-1}_{\max}, r^{-1}_{\min}\Big],
	\label{eq:operator_G}
\end{align}
with \( g(\zeta,\zeta') \) being the Hermitian convolutional kernel:
\begin{align}
	g(\zeta,\zeta') = g(\zeta-\zeta') & = \iint_{\mathcal{P}} \exp\!\Big(-\frac{j2\pi}{\lambda}\frac{\|\mathbf{p}\|^2}{2}(\zeta-\zeta')\Big)d\mathbf{p}, \nonumber \\
            & \quad \quad \quad \quad \quad \quad \forall\, \zeta,\zeta'\in\Big[r^{-1}_{\max}, r^{-1}_{\min}\Big].
	\label{eq:kernel_inverse_distance_no_proj}
\end{align}
\textcolor{black}{Crucially, because the spatial integration over $\mathbf{p} \in \mathcal{P}$ is decoupled from the distance variables, the kernel $g(\zeta,\zeta')$ inherently depends only on the linear difference $(\zeta - \zeta')$. Consequently, it remains a valid Hermitian convolution kernel regardless of the specific shape of the arbitrary bounded aperture $\mathcal{P}$.}

\begin{lemma} \label{lem:kernel_transformation}
	The linear operators \( \mathcal{V} \) and \( \mathcal{G} \) defined in (\ref{eq:operator_V_simplified}) and (\ref{eq:operator_G}), respectively, have the same eigenvalues.
\end{lemma}
\vspace{-5pt}
\noindent \textbf{Proof:} See Appendix A.

\vspace{5pt}

Lemma~\ref{lem:kernel_transformation} indicates that the number of dominant eigenvalues of \( {\mathcal{V}} \) can be deduced by analyzing \( \mathcal{G} \). Moreover, because \( g(\zeta,\zeta') \) is a Hermitian convolution kernel, the number of dominant eigenvalues of \( \mathcal{G} \) can be inferred from Landau's Theorem (or Szego's Theorem) \cite{landau1975szego}:
\begin{align}
	\text{DoF} = \text{Bandwidth}\big(\hat{g}(f)\big) \times L_{\zeta},
	\label{eq:DoF_formula}
\end{align}
where \( \hat{g}(f) = \int_{-\infty}^{\infty} g( \zeta)e^{-j2\pi f  \zeta }dt \) is the Fourier transform of \(g( \zeta )\), \( L_{\zeta} = {r^{-1}_{min}} - {r^{-1}_{max}} \) represents the length of the interval in the inverse-distance domain. Intuitively, the operator \( \mathcal{G} \) in (\ref{eq:operator_G}) with convolution kernel \( g( \zeta ) \) behaves like a linear time-invariant system in which an input signal of duration \( L_{\zeta} \) is transmitted through a channel with frequency response \( g( \zeta ) \). Thus, the DoF of this system corresponds to the number of data symbols that can be transmitted in this channel in duration \(L_{\zeta}\). In our setting, the time domain is replaced by the inverse-distance domain, and the frequency domain becomes the inverse-distance frequency domain.

For simplicity, we rescale \( \hat{g}(f) \) by \( L_{\zeta} \) (i.e., setting \( \xi = L_{\zeta} f \)). Hence, the DoF is simply equal to the bandwidth of \( \tilde{g}(\xi) \) where:
\begin{align}
	\tilde{g}(\xi) = \hat{g}\!\left(\frac{\xi}{L_{\zeta}}\right) = \int_{-\infty}^{\infty} g( \zeta )e^{-j2\pi \xi \frac{ \zeta }{L_{\zeta}}}d \zeta .
	\label{eq:gtilde_definition}
\end{align}
In the next subsection, we derive a closed-form expression for \( \tilde{g}(\xi) \) and examine its bandwidth, thereby characterizing the spatial DoF of the channel.

%% file: sec3_2_Fourier_analysis_v4.tex
\subsection{Eigenvalue Analysis via Fourier Transform Analysis}
\label{sec:eigenvalue_analysis}

The scaled Fourier transform \( \tilde{g}(\xi) \) defined in (\ref{eq:gtilde_definition}) has the following properties shown in Lemma~\ref{lem:fourier_closedform}.

\begin{lemma}
    \label{lem:fourier_closedform}
    Scaled Fourier transform defined in (\ref{eq:gtilde_definition}) of \( g(\Delta  \zeta) \) in (\ref{eq:kernel_inverse_distance_no_proj}) is given by:
    \begin{align}
        \tilde{g}(\xi) = \tilde{g}_0(\xi) * \mathrm{sinc}(2\xi),
    \label{eq:fourier_closed_form}
    \end{align}
    where \(*\) is the convolution operation, \( \mathrm{sinc}(x) = \frac{ \sin(\pi x) }{\pi x} \), and \( \tilde{g}_0(\xi) \) is defined as follows:
    \begin{align}
	\tilde{g}_0(\xi) = \iint_{\mathbf{p} \in \mathcal{P}} \delta\!\Bigl(\xi + \frac{\|\mathbf{p}\|^2}{2\lambda}\Bigl(r^{-1}_{\min}-r^{-1}_{\max}\Bigr)\Bigr)d\mathbf{p},
        \label{eq:tilde_g0}
    \end{align}
    where \( \delta(.) \) is Dirac delta function. The function \( \tilde{g}_0(\xi) \) is strictly band-limited within the following interval:
	\begin{align}
		\Omega \in \left[-\frac{p_{\max}^2}{2\lambda}\Bigl(r^{-1}_{\min}-r^{-1}_{\max}\Bigr),\, -\frac{p_{\min}^2}{2\lambda}\Bigl(r^{-1}_{\min}-r^{-1}_{\max}\Bigr)\right],
        \label{eq:bandlimited_range}
	\end{align}
    and stricly equal to zero outside the interval \( \Omega \).
\end{lemma}
\vspace{-5pt}
\textbf{Proof:} Proof is provided in Appendix B. \textcolor{black}{Note that the Dirac delta function in the definition of $\tilde{g}_0(\xi)$ dictates that its non-zero support occurs exclusively where $\xi = - \frac{\|\mathbf{p}\|^2}{2\lambda}\big(r^{-1}_{\min}-r^{-1}_{\max}\big)$. Because $r_{\max} > r_{\min}$, the inverse-distance term $(r^{-1}_{\min}-r^{-1}_{\max})$ is strictly positive. Consequently, the spatial-frequency variable $\xi$ is strictly negative.}

\vspace{5pt}

\begin{figure}[ht]
    \centering
    \vspace{-20pt}
    \includegraphics[width=7.5cm,height=12.5cm]{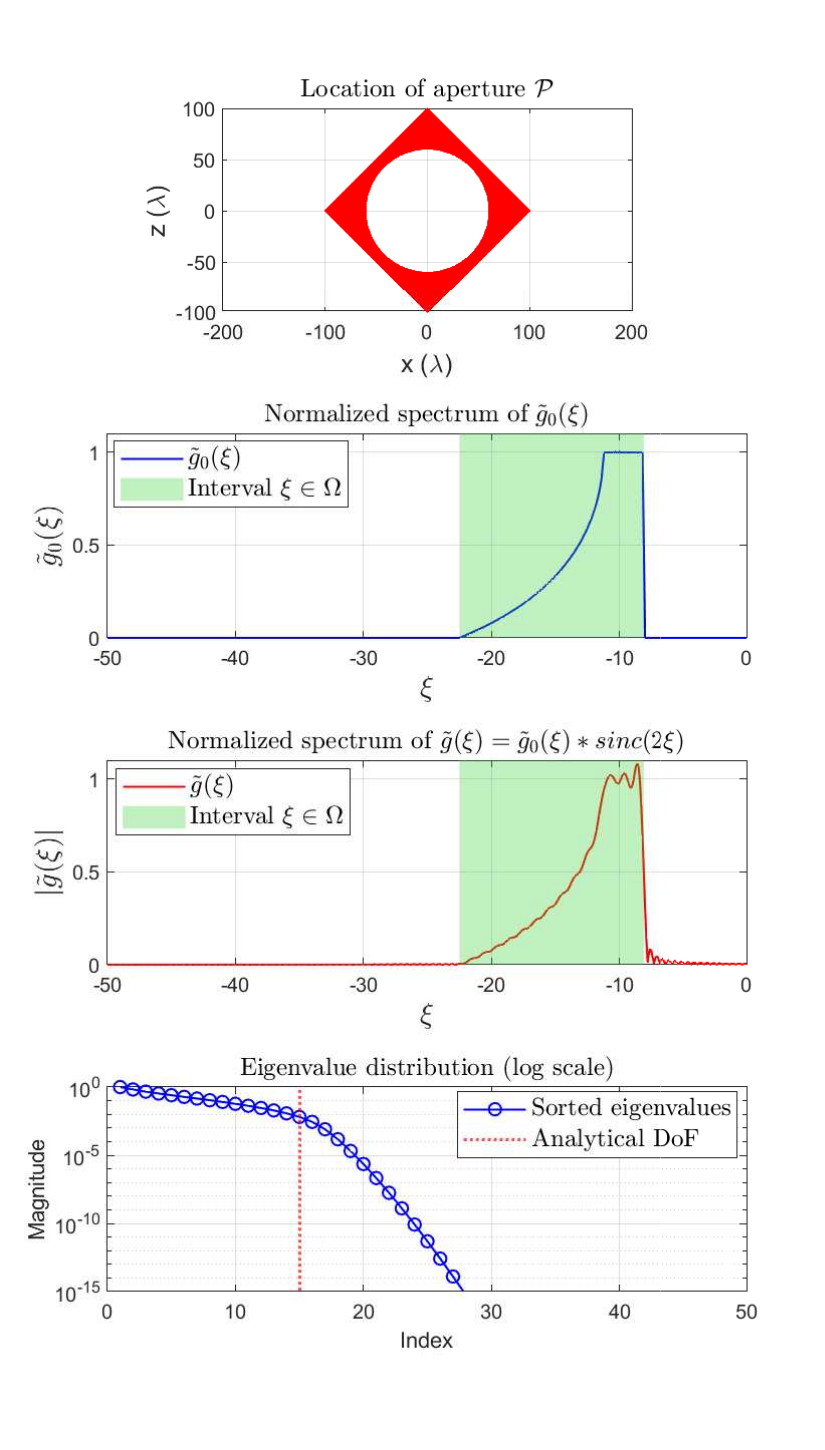}
    \vspace{-30pt}
    \caption{Geometry of the Tx array $\mathcal{P}$ in the xz-plane (top), the corresponding normalized spectra of $\tilde{g}_0(\xi)$ and $\tilde{g}(\xi)$ (middle panels), and eigenvalue distribution (bottom). The squared Tx array spans $[60\lambda, 100\lambda]$ with a circular gap (radius $60\lambda$). The Rx array spans $[200\lambda, 2000\lambda]$, and the analytical DoF ignores the $O(1)$ term.}
    \label{fig:continuous_rectangular_array_FT}
\end{figure}

Lemma~\ref{lem:fourier_closedform} determines the bandwidth of \( \tilde{g}(\xi) \) as follows:
\begin{align}
    \frac{\left(p_{\max}^2 - p_{\min}^2\right)}{2\lambda} \left(r^{-1}_{\min} - r^{-1}_{\max}\right) + O(1),
\end{align}
where the first part is the strict bandwidth of \( \tilde{g}_0(\xi) \) and \( O(1) \) is the additional effective bandwidth due to the sidelobes of the sinc function, whose bandwidth is exactly 1. This reasoning directly leads to the main result stated in Theorem \ref{thm:DoF_infinite}, which quantifies the DoF of the operator \( \mathcal{G} \) and hence \( \mathcal{V} \). To illustrate Lemma~\ref{lem:fourier_closedform} and Theorem~\ref{thm:DoF_infinite}, we plot the spectrum of \( \tilde{g}(\xi) \) and \( \tilde{g}_0(\xi) \) and the eigenvalue distribution for a squared array \( \mathcal{P} \) with circular gap in the center in Fig.~\ref{fig:continuous_rectangular_array_FT}.


\begin{theorem}[Distance-Domain DoF Expression for Broadside Case]
    \label{thm:DoF_infinite}
    Consider the LoS channel between an arbitrary single-piece Tx array and a Rx linear array, where both arrays consist of an infinite number of elements with infinitesimal spacing. For the Rx array spanning over \( [r_{\min},r_{\max}] \), and the single-piece Tx array bounded by \(p_{\min}\) and \(p_{\max}\), the distance-domain DoF (i.e., the number of significant eigenvalues) is given by
    \begin{align}
	\text{DoF} = \frac{\left(p_{\max}^2 - p_{\min}^2\right)}{2\lambda} \left(r^{-1}_{\min} - r^{-1}_{\max}\right) + O(1),
    \label{eq:DoF_infinite_final}
    \end{align}
    \textcolor{black}{where \(O(1)\) denotes a bounded residual term that inherently depends on the specific aperture geometry and the chosen eigenvalue threshold.}
\end{theorem}

\textcolor{black}{Note that for scenarios with large distance spans or exceedingly large antenna arrays, the main term grows significantly, making the \(O(1)\) term practically negligible. In such regimes, the effective DoF is primarily dominated by the main analytical term, and hence, can be simplified as follows:
\begin{align}
    \text{DoF} \approx \left\lceil \frac{\left(p_{\max}^2 - p_{\min}^2\right)}{2\lambda} \left(r^{-1}_{\min} - r^{-1}_{\max}\right) \right\rceil,
    \label{eqn:simplified_analytical_DoF}
\end{align}
which depends strictly on the aperture's extreme edges rather than its detailed geometry or the specific eigenvalue threshold.}

\textcolor{black}{To explicitly validate this and quantify the \(O(1)\) behavior, we simulate the eigenvalue distribution for three distinct aperture shapes sharing the same radial extremes (\(p_{\min} = 60\lambda, p_{\max} = 100\lambda\)) and observation span (\(r_{\min}=200\lambda, r_{\max}=2000\lambda\)). For the feasibility of numerical evaluation, we employ discrete arrays with half-wavelength spacing for both the transmit aperture and the observation array. For this configuration, the main analytical term evaluates to \(\left\lceil\frac{\left(p_{\max}^2 - p_{\min}^2\right)}{2\lambda} \left(r^{-1}_{\min} - r^{-1}_{\max}\right) \right\rceil = 15\). The \(O(1)\) term can then be numerically extracted as the residual difference between the simulated effective DoF and this main term. We demonstrate that \(O(1)\) fluctuates only slightly across different aperture shapes and thresholds, remaining highly insignificant compared to the main term. For example, applying an effective DoF threshold of \(\epsilon=10^{-1}\), the \(O(1)\) term evaluates to \(0\), \(1\), and \(1\) for the rectangular, circular, and linear arrays, respectively. Even when applying a more stringent threshold of \(\epsilon = 10^{-3}\) to capture a wider transition band, the \(O(1)\) term only marginally increases to \(1\), \(2\), and \(2\).}

\textcolor{black}{It is important to emphasize that despite using continuous arrays as the analytical framework, our proposed analytical DoF accurately predicts the performance of discrete arrays with half-wavelength spacing. To demonstrate this, we numerically evaluated the eigenvalue distributions assuming half-wavelength spacing for different aperture shapes sharing the same extreme edges in Fig.~\ref{fig:different_shapes}, as well as for apertures with different extreme edges in Fig.~\ref{fig:different_sizes}. The results confirm that the numerical DoF obtained from discrete half-wavelength arrays tightly matches our continuous analytical limits.}

Furthermore, if we define
\(\dot{p}_{\max} \triangleq p_{\max}/\lambda\), \(\dot{p}_{\min} \triangleq p_{\min}/\lambda\),
\(\dot{r}_{\max} \triangleq r_{\max}/\lambda\), and \(\dot{r}_{\min} \triangleq r_{\min}/\lambda\).
Then the distance-domain DoF depends only on these normalized extreme edges:
\begin{align}
    \text{DoF} \approx
    \frac{\big(\dot{p}_{\max}^2 - \dot{p}_{\min}^2\big)\big(\dot{r}_{\min}^{-1} - \dot{r}_{\max}^{-1}\big)}{2}
    + O(1),
    \label{eq:DoF_infinite_normalized}
\end{align}
Accordingly, in what follows we numerically evaluate DoF using normalized edge coordinates rather than specifying a particular carrier frequency or physical aperture.

\subsection{Comparison with Other DoF Expressions in The Literature}

\subsubsection{\textbf{Comparison with Angular-Domain DoF Expressions}}
In conventional point-to-point (P2P) MIMO setups, the Rx aperture primarily spans the \emph{angular} extent of the Tx aperture. Consequently, existing angular-domain DoF expressions typically take the following form~\cite{torkildson2011indoor,pizzo2022landau,miller2000communicating,xie2023performance,pu2014effects,ruiz2023degrees,do2023parabolic,xu2017degrees,yuan2021electromagnetic,xu2023exploiting,guo2025impact}:
\begin{align}
\mathrm{DoF}
\;=\;
\frac{\alpha\,A_t A_r}{(\lambda D)^n} \;+\;  O(1),
\label{eqn:DoF_expression_literature}
\end{align}
where \(A_t\) and \(A_r\) denote the physical areas of the Tx and Rx arrays, \(D\) is the Tx–Rx separation (implicitly treated as a single, uniform distance), \(n\) depends on the aperture dimensionality, and \(\alpha\) is a coefficient determined by the array orientations. More elaborate variants appear in~\cite{wang2025analytical}. Our distance-domain result derived in~\eqref{eq:DoF_infinite_final} fundamentally diverges from this standard model in two key aspects:

\begin{figure}[ht]
    \centering
    \includegraphics[width=8.5cm]{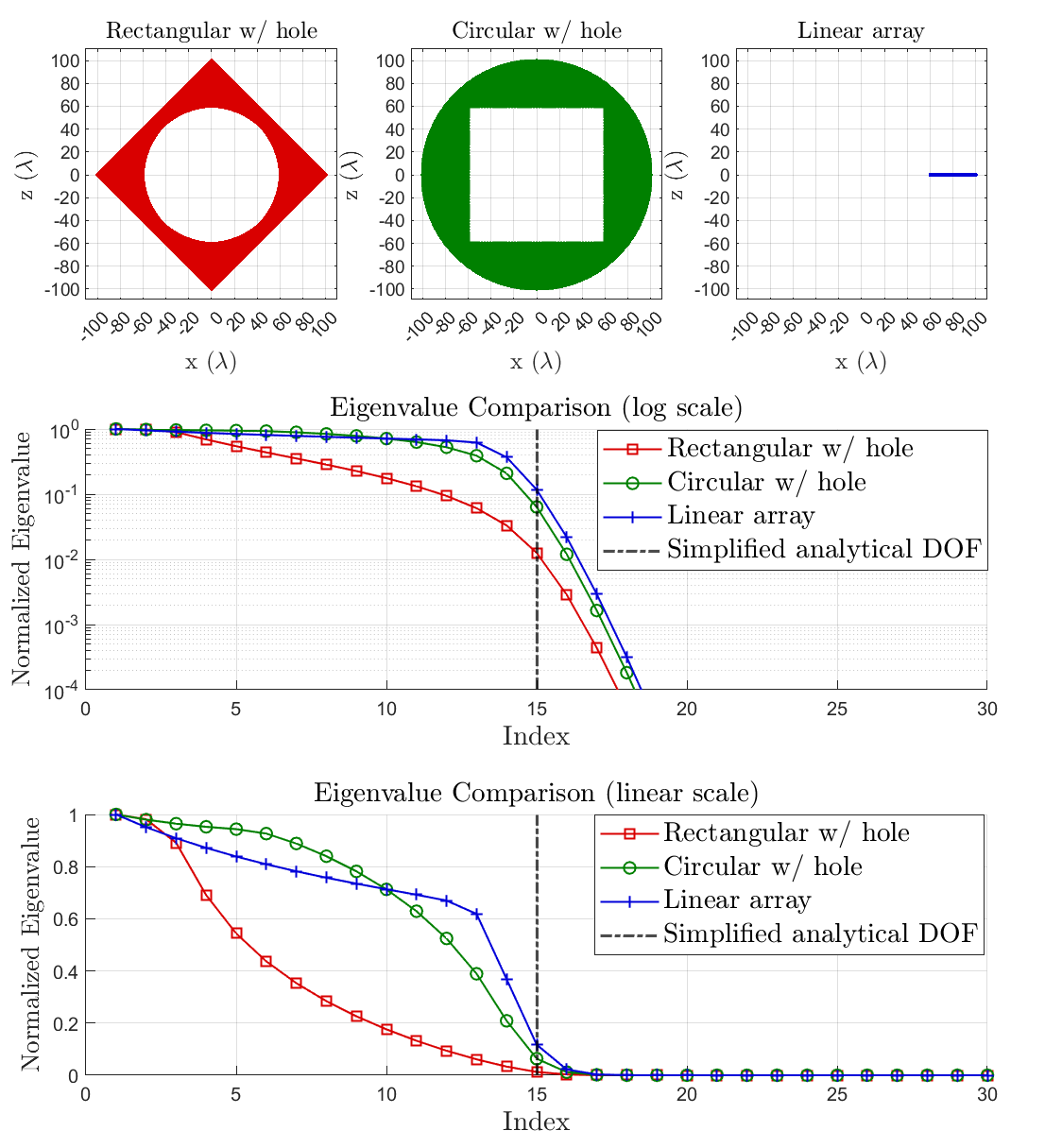}
    \caption{Three different Tx array geometries in the xz-plane (top panels) and their corresponding normalized eigenvalue distributions in log scale (middle panel) and linear scale (bottom panel). All three shapes share the same radial bounds ($p_{\min}=60\lambda$ and $p_{\max}=100\lambda$). The analytical DoF is obtained via (\ref{eqn:simplified_analytical_DoF}).}
    \label{fig:different_shapes}
\end{figure}

\begin{figure}[ht]
    \centering
    \includegraphics[width=9cm]{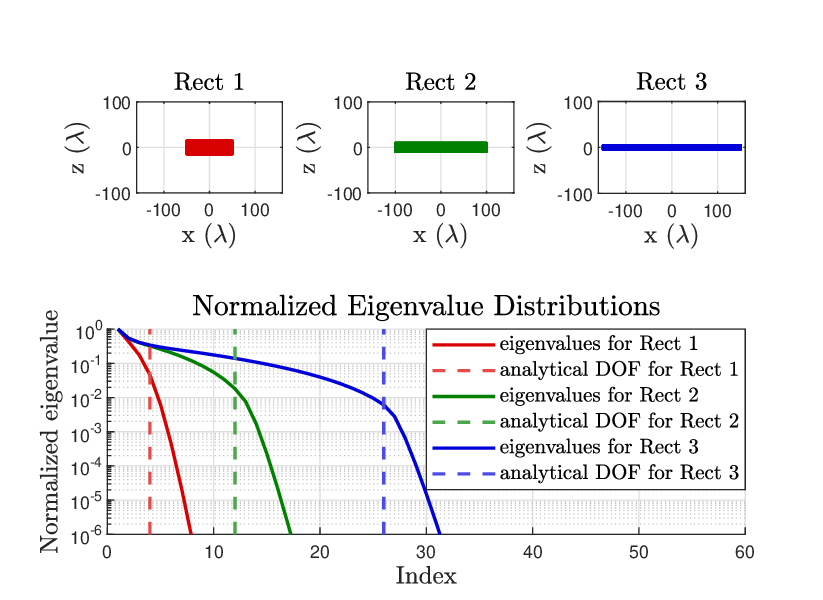}
    \caption{Geometries of three rectangular Tx arrays sharing the exact same physical area (top panels) and their corresponding normalized eigenvalue distributions (bottom panel). The array dimensions are $50\lambda \times 30\lambda$ (Array 1), $100\lambda \times 15\lambda$ (Array 2), and $150\lambda \times 10\lambda$ (Array 3). The broadside linear Rx array $\mathcal{Q}$ spans from $r_{\min}=400\lambda$ to $r_{\max}=4000\lambda$.}
    \label{fig:different_sizes}
\end{figure}

\textcolor{black}{\emph{\textbf{1) Scaling with Tx Aperture Geometry:}} 
Unlike angular-domain DoF, which scales monotonically with the physical array area (e.g., linearly with ULA length or quadratically with URA side length), our distance-domain DoF depends solely on the difference of the squared radial extremes, \(\big(p_{\max}^{2}-p_{\min}^{2}\big)\). As demonstrated in Fig.~\ref{fig:different_shapes}, arrays with vastly different detailed shapes and physical areas yield nearly identical distance-domain DoF provided they share the same radial boundaries \((p_{\max}, p_{\min})\). Consequently, for a fixed aperture area, strategically elongating the array to maximize \(p_{\max}\) is far more effective for increasing the DoF than uniformly enlarging the area. This geometric scaling advantage is explicitly verified in Fig.~\ref{fig:different_sizes}, which compares three rectangular arrays with identical physical areas but varying elongations.}

\textcolor{black}{\emph{\textbf{2) Scaling with Rx Span and Distance Spread:}} 
Angular-domain models implicitly assume a narrow spatial span (\(r_{\min} \approx r_{\max} \approx D\)), causing the DoF to scale simply with the receive area and \(D^{-n}\). In stark contrast, our framework accurately models systems with exceptionally large distance spreads (\(r_{\max} \gg r_{\min}\)). The distance-domain DoF scales with the \emph{inverse-distance span}, \(\big(\frac{1}{r_{\min}}-\frac{1}{r_{\max}}\big)\), rather than the absolute physical length of the Rx array. A profound physical consequence of this scaling is that as the receive observation region extends toward infinity (\(r_{\max}\to\infty\)), the DoF does not exhibit unphysical, unbounded growth; instead, it safely saturates to a finite limit bounded by \(1/r_{\min}\).}

\vspace{5pt}

\textcolor{black}{\subsubsection{\textbf{Comparison with Distance-Domain DoF in the Literature}}
We compare our framework with \cite{ding2022degrees}, which analyzes two perpendicular ULAs---the geometry closest to our broadside setting. By mapping their geometric variables (center-to-center distance $D$, and ULA lengths $L_t, L_r$) to our radial notation ($D=(r_{\max}+r_{\min})/{2}$, $L_r=r_{\max}-r_{\min}$, and $p_{\max}=L_t/2$), their closed-form DoF expression \cite[Eqs.~(39),(33)]{ding2022degrees} translates directly to:
\begin{align}
\mathrm{DoF}_{\text{\cite{ding2022degrees}}}
= \frac{r_{\max}}{2\lambda}
\!\left(
\frac{r_{\max}}{\sqrt{r_{\max}^{2}+p_{\max}^{2}}}
-\frac{r_{\min}}{\sqrt{r_{\min}^{2}+p_{\max}^{2}}}
\right).
\label{eqn:DoF_ULA_ULA_ref_equiv}
\end{align}}

\begin{figure}[ht]
    \centering
    \includegraphics[width=9cm, height=5cm]{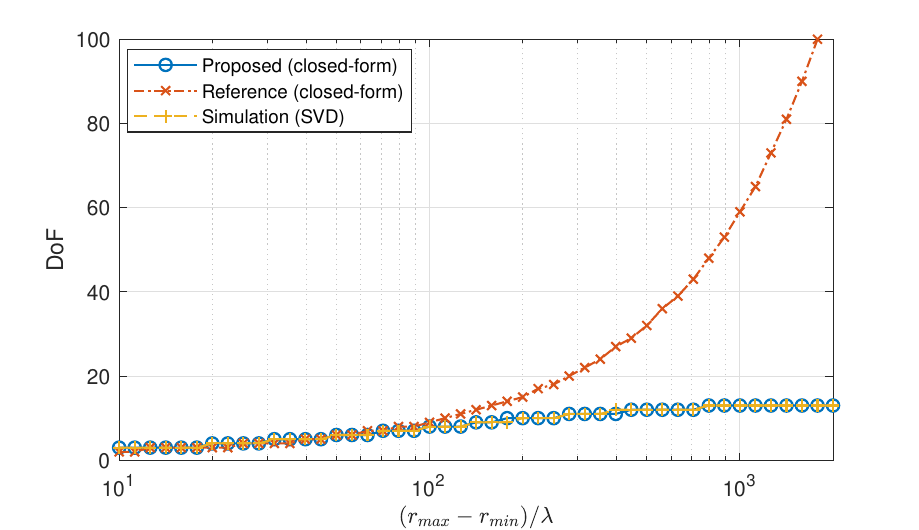}
    \caption{Distance-domain DoF vs. length of Rx array obtained via: the proposed analytical DoF, the expression from \cite{ding2022degrees}, and SVD simulations (threshold \(\epsilon=0.1\sigma_{\max}\)). The Tx array is a ULA of length \(100\lambda\), \(r_{\min}=100\lambda\), and the Rx array length (\(r_{\max}-r_{\min}\)) ranges between \([10\lambda, 2000\lambda]\).}
    \label{fig:proposed_paper_vs_JSAC}
\end{figure}

\textcolor{black}{As illustrated in Fig.~\ref{fig:proposed_paper_vs_JSAC}, the validity of \eqref{eqn:DoF_ULA_ULA_ref_equiv} depends heavily on the distance span. For \emph{narrow} distance spans ($D \gg L_r$, or $r_{\max} \approx r_{\min}$), both \eqref{eqn:DoF_ULA_ULA_ref_equiv} and our proposed expression in \eqref{eq:DoF_infinite_final} reduce to the same approximation, $\mathrm{DoF} \approx \frac{(r_{\max}-r_{\min})p_{\max}^{2}}{2\lambda D^{2}}$, tightly matching the numerical simulations.}

\textcolor{black}{However, a critical divergence occurs for \emph{large} distance spans. As \(r_{\max}\to\infty\), the expression from \cite{ding2022degrees} grows without bound. In stark contrast, our proposed DoF remains physically finite, correctly saturating to $\frac{p_{\max}^{2}-p_{\min}^{2}}{2\lambda\,r_{\min}}$. Thus, when users or observation regions are spread over a wide range of distances, our framework captures the correct bounded scaling, whereas the prior model in \cite{ding2022degrees} severely overestimates the available DoF.}

\textcolor{black}{\textbf{Remark:} To summarize, our framework overcomes the fundamental limitations of both existing angular-domain and distance-domain DoF models in \cite{ding2022degrees}. While conventional angular-domain models scale heavily with aperture area, and prior distance-domain works like \cite{ding2022degrees} are restricted to specific array geometries and narrow distance spans (\(r_{\min}\!\approx\! r_{\max}\)), our derived expression in \eqref{eq:DoF_infinite_final} offers a universal solution. It applies to arbitrary two-dimensional transmit apertures by depending solely on radial extremes, and it remains strictly well-posed for unbounded distance spreads (\(r_{\max} \rightarrow \infty\)). Consequently, our model uniquely captures the spatial multiplexing capabilities of realistic near-field MU-MIMO networks where both the base station aperture shape and user distance separations can be highly generalized.}

\vspace{5pt}
\subsubsection{\textbf{Non-Uniformity of the Eigenvalue Spectrum}}

A common (often implicit) assumption in DoF estimates is a \emph{\textbf{nearly flat}} plateau of dominant eigenvalues followed by rapid decay. This is reasonable when the channel kernel is (approximately) convolutional and the Fourier transform of its kernel is close to an in-band indicator over a passband \(\Omega\) (flat in-band)~\cite{franceschetti2015landau}:
\[
\widetilde h(\boldsymbol\omega)\;\approx\;
\begin{cases}
1, & \boldsymbol\omega\in\Omega,\\
0, & \text{otherwise}.
\end{cases}
\]

In our distance-domain setting, the relevant spectral density \(\tilde g(\xi)\) is \emph{\textbf{not necessarily flat}} over its main lobe (see Fig.~\ref{fig:continuous_rectangular_array_FT}). As a result, the leading eigenvalues are generally \emph{non-uniform}: they taper gradually before the fast exponential-decay region set by the analytical DoF threshold. The taper profile also depends on the Tx-aperture shape (Fig.~\ref{fig:different_shapes}), since different shapes induce different \(\tilde g(\xi)\).

\emph{Implication}: 
Effective-DoF method such as \cite{muharemovic2008antenna,guo2025impact,yuan2021electromagnetic,wang2025analytical,bai2025dynamic}:
\begin{align}
    \text{DoF}_{\text{eff}}
    = \frac{(\text{tr}\,\mathbf{V})^2}{\text{tr}\,(\mathbf{V}^{\text{H}}\mathbf{V})}
    = \frac{\big(\sum_{k}\lambda_k\big)^2}{\sum_{k}\lambda_k^2},
    \qquad \mathbf{V}=\mathbf{H}^{\text{H}}\mathbf{H},
    \label{eq:eff_rank_proxy}
\end{align}
are most accurate when the leading eigenvalues are nearly equal (\(\lambda_1\!\approx\!\cdots\!\approx\!\lambda_N\!\gg\!\lambda_{N+1}\)). With our non-uniform eigenvalue spectra, \eqref{eq:eff_rank_proxy} can bias the DoF (often downward). By contrast, the bandwidth-based count (via Landau's Theorem) tracks the {bandwidth} of \(\tilde g(\xi)\), which is insensitive to detailed shape given the same radial extremes \((p_{\min},p_{\max})\). This yields more correct leading-order DoF as shown in Fig.~\ref{fig:different_shapes}.

%% file: sec4_nonBroadside_v4.tex
\section{Extension to Non-Broadside Receive Array}
\label{sec:non_broadside_case}

\subsection{Transformation to Broadside Case via Projection}

\begin{figure}[ht]
    \centering
    \includegraphics[width=0.5\linewidth]{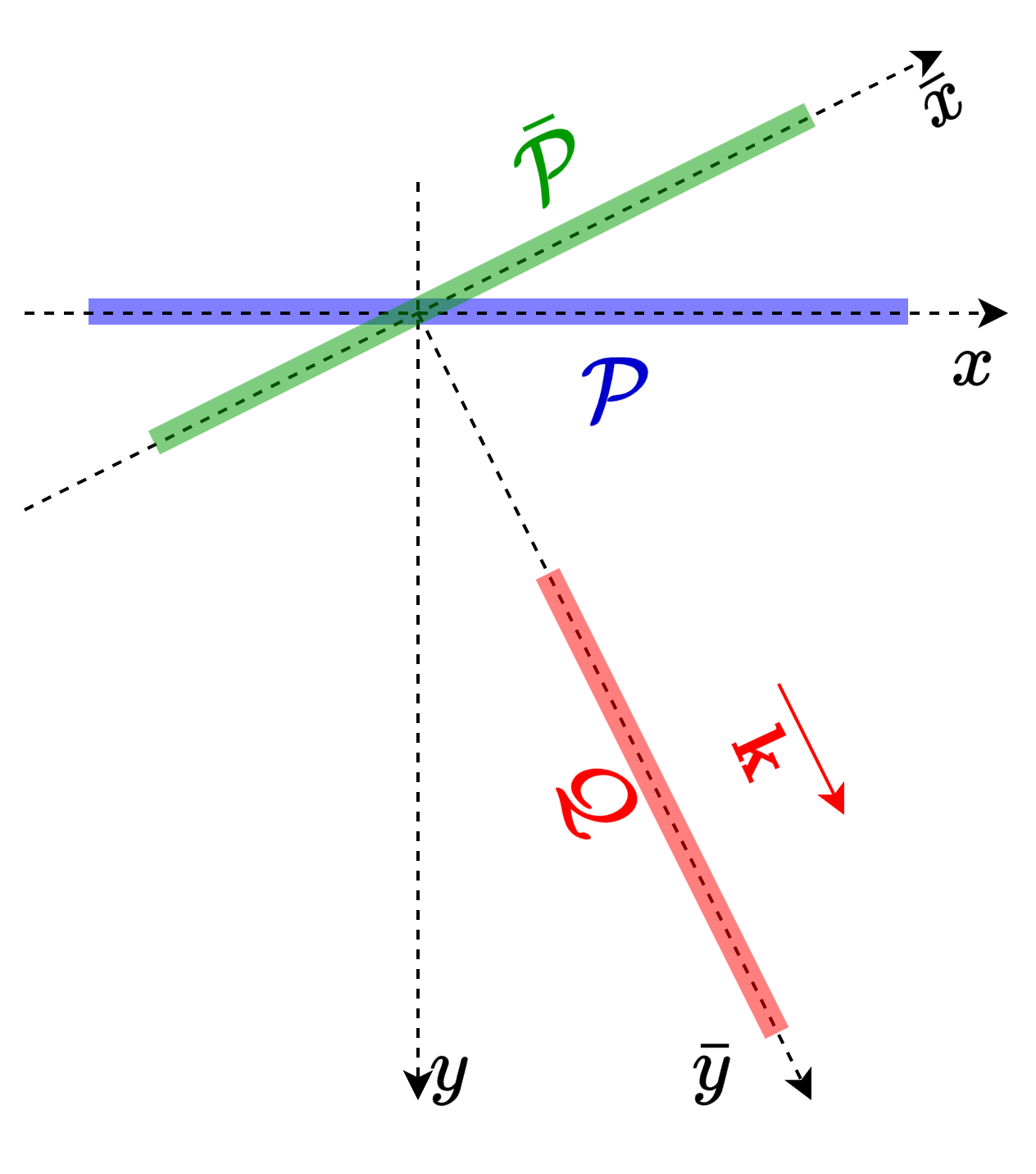}
    \caption{Projected Tx array \(\bar{\mathcal{P}}\): projection of Tx array \({\mathcal{P}}\) onto the plane orthogonal to the direction vector \(\mathbf{k}\) (For simplicity of illustration, we limit to only the x- and y-axes).}
    \label{fig:projection2}
\end{figure}

\textcolor{black}{To analyze the non-broadside configuration efficiently, we project the Tx array \(\mathcal{P}\) onto the plane orthogonal to the Rx array's direction vector \(\mathbf{k}\). Using the projection matrix \( \mathbf{I}_3 - \mathbf{k}\mathbf{k}^T \), the \emph{projected Tx array} is defined concisely as:
\begin{equation}
    \bar{\mathcal{P}} = \left\{ \bar{\mathbf{p}} = \left( \mathbf{I}_3 - \mathbf{k}\mathbf{k}^T \right)\mathbf{p} \;\middle|\; \mathbf{p} \in \mathcal{P} \right\},
    \label{eq:projected_aperture}
\end{equation}
where any projected point \( \bar{\mathbf{p}} \in \bar{\mathcal{P}} \) satisfies the geometric condition \(\|\bar{\mathbf{p}}\|^2 = \|\mathbf{p}\|^2 - (\mathbf{k}^T \mathbf{p})^2\).
Substituting this relationship into the non-broadside channel model, the corresponding kernel function can be rewritten as an integral over the projected aperture \(\bar{\mathcal{P}}\):
\begin{align}
    {v}(r,r') &\approx \frac{1}{rr'}\iint_{\bar{\mathcal{P}}} \exp\!\Bigg(-\frac{j2\pi}{\lambda}\frac{\|\bar{\mathbf{p}}\|^2}{2}\Big(\frac{1}{r}-\frac{1}{r'}\Big)\Bigg)d\bar{\mathbf{p}}.
    \label{eq:kernel_nonbroadside}
\end{align}
Crucially, this kernel function is exactly similar to the broadside case derived in (\ref{eq:kernel_broadside}), with the original coordinates \(\mathbf{p}\) and aperture \(\mathcal{P}\) simply replaced by their projected counterparts \(\bar{\mathbf{p}}\) and \(\bar{\mathcal{P}}\).}

Because the mathematical structure is identical, we can directly apply Theorem~1 to this equivalent broadside system. The distance-domain DoF is obtained by simply substituting the original radial boundaries in \eqref{eq:DoF_infinite_final} with the extreme radial distances of the projected array (\(\bar{p}_{\min}\) and \(\bar{p}_{\max}\)):
\begin{align}
    \text{DoF} \approx \frac{\left(\bar{p}_{\max}^2 - \bar{p}_{\min}^2\right)}{2\lambda} \left(\frac{1}{r_{\min}} - \frac{1}{r_{\max}}\right) + O(1).
    \label{eq:DoF_infinite_nonbroadside_final}
\end{align}

\begin{figure}[ht]
    \centering
    \includegraphics[width=8cm,height=4cm]{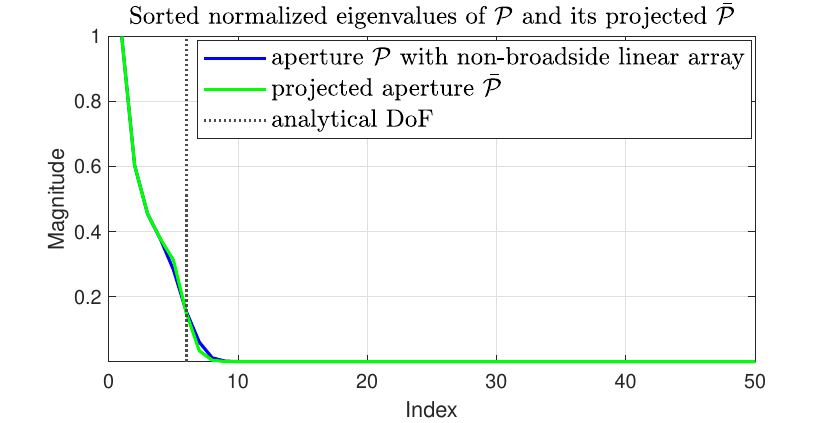}
    \caption{Eigenvalue distribution comparison: 1) Tx array \( \mathcal{P} \) and non-broadside Rx array \( \mathcal{Q} \), versus 2) Projected Tx array \( \bar{\mathcal{P}} \) and the same Rx array \( \mathcal{Q} \). (\( \mathcal{P} \) is a \( 150\lambda \times 50\lambda \) rectangular array located at the origin; the linear array \( \mathcal{Q} \) has \( \phi=\frac{\pi}{3}, \theta = \frac{\pi}{2} \), with \( r_{\min} = 400\lambda \) and \( r_{\max} = 4000\lambda \)).}
    \label{fig:eigenvalues_nonbroadside_vs_projection}
\end{figure}

\begin{figure}[ht]
    \centering
    \includegraphics[width=8.5cm,height=7.5cm]{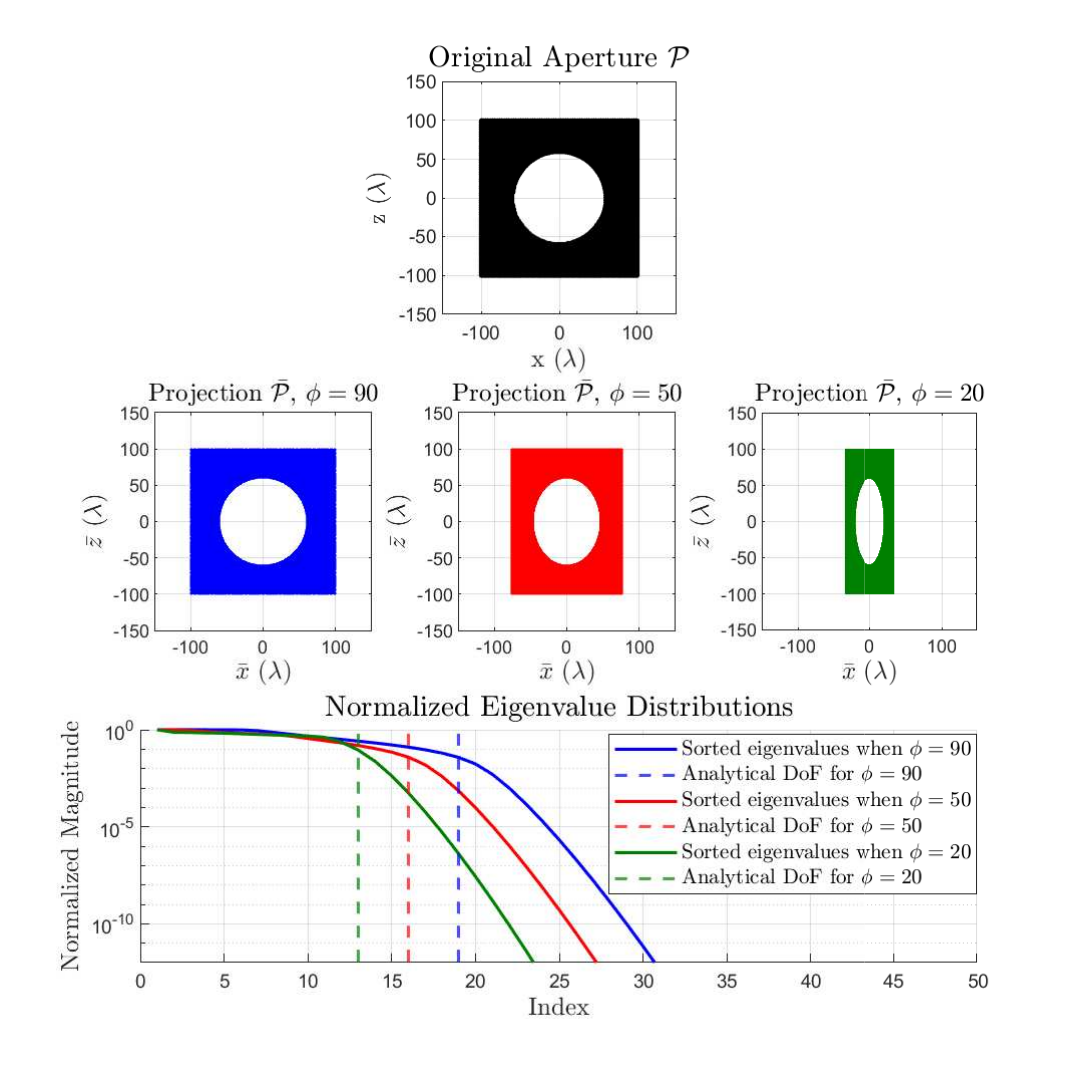}
    \caption{Original Tx array aperture (top panel) and its projections for different Rx array angles $\phi$ (middle panels) and the corresponding eigenvalue distributions (bottom panel) of the corresponding channels. The Rx array spans from $r_{\min}=400\lambda$ to $r_{\max}=4000\lambda$.}
\label{fig:different_angles}
\end{figure}

As verified by the eigenvalue distributions in Fig.~\ref{fig:eigenvalues_nonbroadside_vs_projection}, this projected broadside model perfectly characterizes the true non-broadside channel. Physically, the distance-domain DoF in the non-broadside configuration is fundamentally lower than in the broadside case. As illustrated in Fig.~\ref{fig:different_angles}, when the Rx array $\mathcal{Q}$ is oriented further away from the broadside direction, the effective projected Tx array $\bar{\mathcal{P}}$ shrinks. This geometric reduction directly decreases the extreme radial values $\bar{p}_{\max}$ and $\bar{p}_{\min}$, which in turn restricts the available distance-domain DoF.

\subsection{Approximation Accuracy for Distance-Domain DoF in the Non-Broadside Case}
\label{subsec:approx_accuracy_nbb}

Our analytical DoF expression relies on two standard simplifications: a Fresnel phase approximation and a uniform-amplitude model (which neglects polarization and element-wise distance variations). To assess the robustness of these assumptions, we compare three channel models: (i) our baseline \emph{Fresnel phase \& uniform amplitude}; (ii) \emph{exact phase \& uniform amplitude}; and (iii) \emph{exact phase \& exact amplitude}, which incorporates full polarization and distance-dependent path loss \cite{lu2021communicating}.

For a propagation direction $\mathbf{k}_k$ and a Tx element at $\mathbf{p}\!\in\!\mathcal{P}$, the worst-case residual phase error introduced by the Fresnel expansion over the Tx aperture is bounded by:
\begin{equation}
  e_{\text{phase}}
  \;\le\;
  \frac{2\pi}{\lambda}\,
  \frac{\bigl(\mathbf{k}_k^{\mathsf T}\mathbf{p}_{\max}\bigr)\Bigl(\|\mathbf{p}_{\max}\|^2 - \bigl(\mathbf{k}_k^{\mathsf T}\mathbf{p}_{\max}\bigr)^2\Bigr)}
       {2\,r_{\min}^2}.
  \label{eq:fresnel_error_bound}
\end{equation}
The \emph{Fresnel distance} is defined as the minimum range $r_{\min}$ at which this maximum error falls below a prescribed threshold (e.g., $\pi/4$).

\textcolor{black}{We numerically compute the eigenvalue spectra for these three models using a $160\lambda\times160\lambda$ square Tx array (where $p_{\max} = 80\sqrt{2}\lambda$). The $z$-polarized Rx array is positioned at an azimuth $\phi=75^\circ$ with a fixed $r_{\max}=4000\lambda$. For this specific configuration, the Fresnel distance for a $\pi/4$ maximum phase error evaluates to approximately $1060\lambda$.}

\begin{figure}[ht]
  \centering
  \vspace{-6pt}
  \includegraphics[width=9cm]{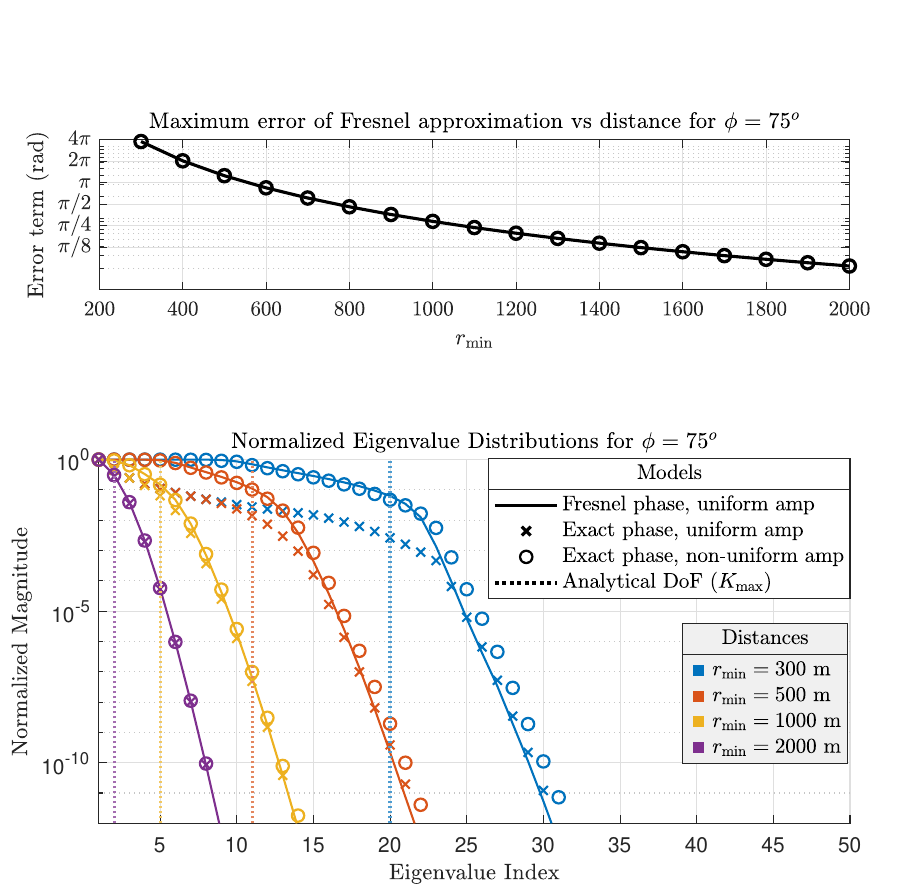}
  \caption{(Top) Maximum Fresnel phase error versus $r_{\min}$; (Bottom) normalized eigenvalue distributions for the three channel models compared with the analytical DoF, with $\phi=75^\circ$ and various $r_{\min}/\lambda$. The Tx array spans $[-100\lambda,100\lambda]$.}
  \label{fig:error_Fresnel_approximation_vs_distance}
\end{figure}

\textcolor{black}{As illustrated in Fig.~\ref{fig:error_Fresnel_approximation_vs_distance}, when the Tx--Rx separation meets or exceeds the Fresnel distance (e.g., $r_{\min} \ge 1000\lambda$), the spectra of all three models perfectly coincide with our analytical prediction. This confirms that amplitude non-idealities and phase errors are utterly negligible in this regime. More importantly, even when $r_{\min}$ falls deep within the Fresnel region (e.g., $r_{\min} \le 500\lambda$), where minor spectral deviations naturally arise, the \emph{eigenvalue cutoff index} remains highly consistent across all models---deviating by less than $10\%$ from our analytical prediction. This confirms that our derived closed-form expression is highly robust against amplitude variations, successfully predicting the distance-domain DoF even when strict Fresnel conditions are relaxed.}

%% file: sec5_modular_array_v4.tex
\section{Extensions to Distance-Domain Degrees of Freedom with Modular Arrays}
\label{sec:modular_aperture}

In this section, we extend our fundamental analysis to a more generalized architecture where the Tx array \( \mathcal{P} \) consists of multiple non-overlapping sub-arrays.

\begin{figure}[ht]
    \centering
    \includegraphics[width=6cm]{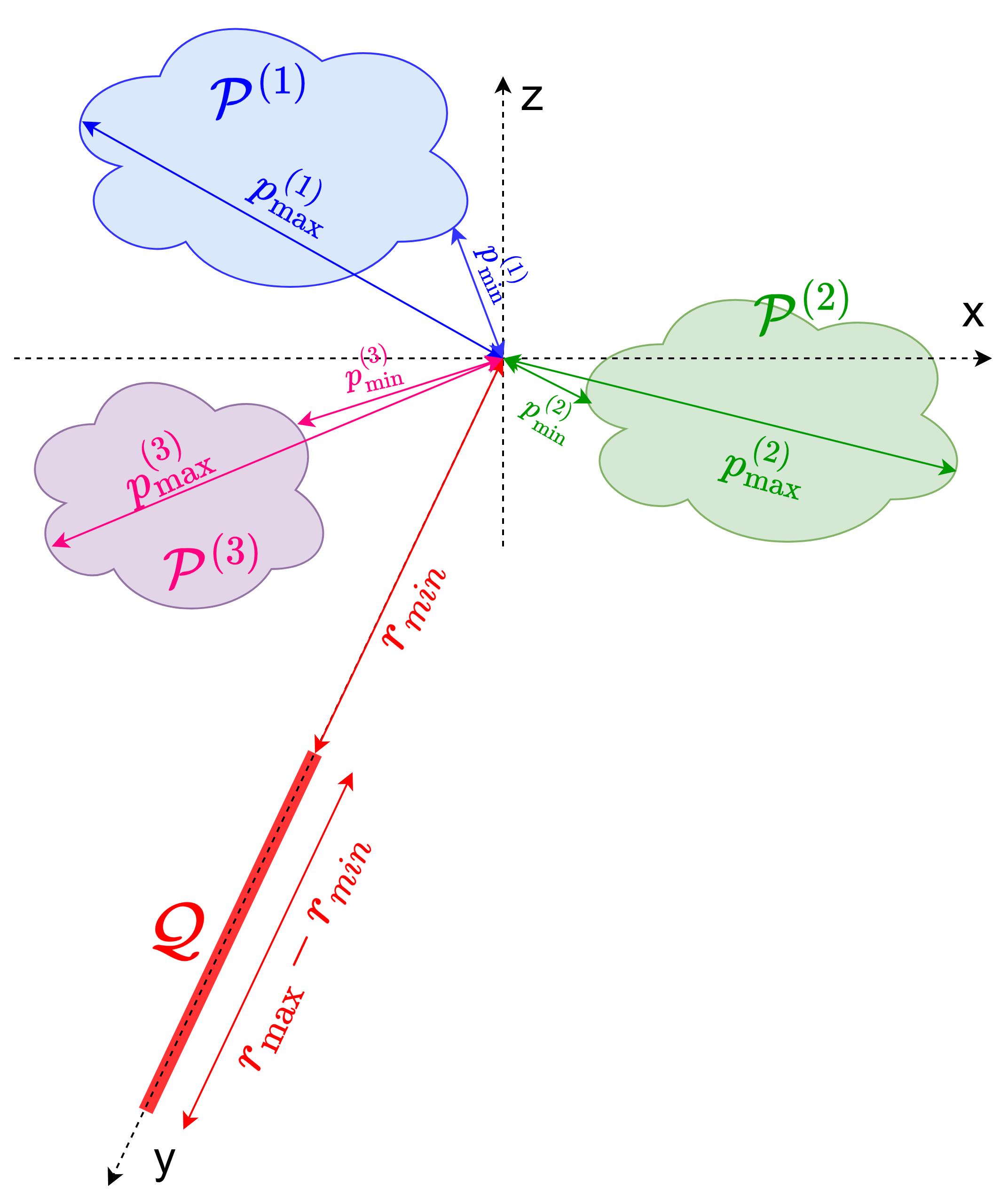}
    \caption{Geometric setup of the Tx array \(\mathcal{P}\) comprising multiple sub-arrays and the broadside Rx linear array \(\mathcal{Q}\).}
    \label{fig:broadside_collinear_elements_3}
\end{figure}

\subsection{Analytical Framework for Spatial DoF of Modular Arrays}

Consider a Tx array partitioned into \(N\) disjoint continuous sub-arrays, \(\mathcal{P} = \bigcup_{n=1}^N \mathcal{P}^{(n)}\). The extreme radial distances for each individual sub-array \( \mathcal{P}^{(n)} \) are defined as:
\begin{align}
    p^{(n)}_{\min} = \min_{ \mathbf{p} \in \mathcal{P}^{(n)} } \|\mathbf{p}\|, \quad p^{(n)}_{\max} = \max_{ \mathbf{p} \in \mathcal{P}^{(n)} } \|\mathbf{p}\|.
    \label{eq:extremes_subarray}
\end{align}

Because the Fourier transform (cf. Lemma~\ref{lem:fourier_closedform}) is a linear operator, the total spatial spectrum \( \tilde{g}(\xi) \) is simply the superposition of the spectra from each sub-array: 
\begin{align}
    \tilde{g}(\xi) = \sum_{n=1}^N \tilde{g}^{(n)}(\xi),
\end{align}
Consequently, based on Theorem~1, the overall main-lobe support \( \Omega \) of the modular array is the direct union of the individual sub-intervals:
\begin{align}
    \Omega = \bigcup_{n=1}^N \Omega^{(n)},
\end{align}
where each sub-interval is bounded by its respective radial extremes:
\begin{align}
    \Omega^{(n)} = \left[ -\frac{(p^{(n)}_{\max})^2(r_{\min}^{-1} - r_{\max}^{-1})}{2\lambda}, \; -\frac{(p^{(n)}_{\min})^2(r_{\min}^{-1} - r_{\max}^{-1})}{2\lambda} \right].
    \label{eqn:interval_subarray}
\end{align}

The total effective bandwidth \(\mathcal{B}(\Omega)\), which directly dictates the distance-domain DoF, is the Lebesgue measure of this combined support. Crucially, \(\mathcal{B}(\Omega)\) is determined entirely by how the radial intervals \([p^{(n)}_{\min},\,p^{(n)}_{\max}]\) overlap across the \(N\) sub-apertures.

\begin{figure}[ht]
    \centering
    \includegraphics[width=5cm]{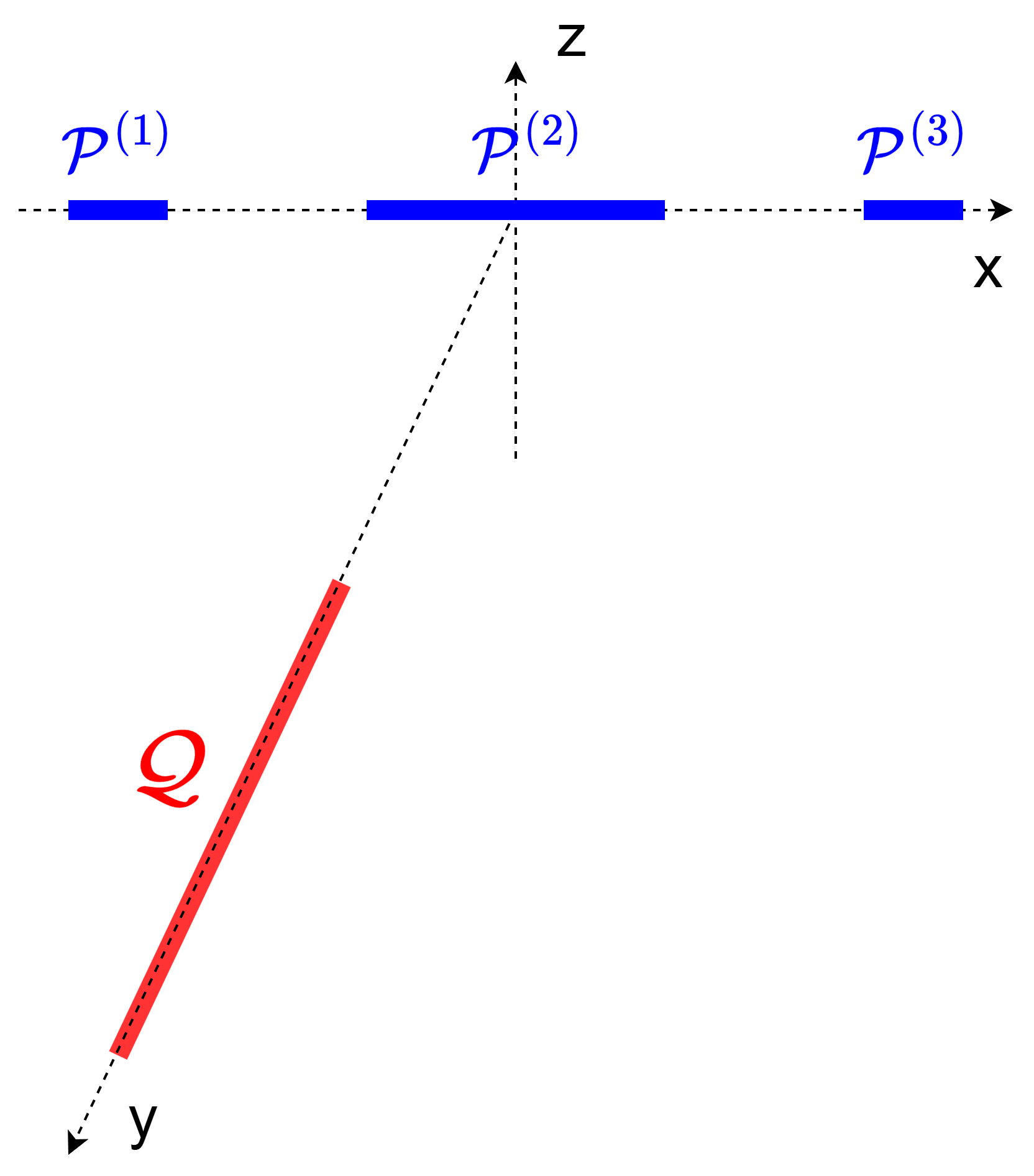}
    \caption{Example: Symmetric Tx array \(\mathcal{P}\) composed of three sub-linear arrays.}
    \label{fig:modular_ULA_broadside_collinear_elements}
\end{figure}

\paragraph*{Example (Three-Subarray ULA)} 
To illustrate this spatial redundancy, consider the symmetric modular ULA composed of three sub-arrays depicted in Fig.~\ref{fig:modular_ULA_broadside_collinear_elements}. Because the two outer sub-arrays (\(\mathcal{P}^{(1)}\) and \(\mathcal{P}^{(3)}\)) are symmetric with respect to the origin, they share the exact same radial extremes. Thus, their spectral supports perfectly overlap: \(\Omega^{(1)} = \Omega^{(3)}\). Assuming the central sub-array \(\mathcal{P}^{(2)}\) occupies a distinctly smaller radial span, it remains disjoint from the outer supports. The total effective bandwidth evaluates to:
\begin{align}
    \mathcal{B}(\Omega)
    \;=\;
    \mathcal{B}\big(\Omega^{(1)}\big)
    \;+\;
    \mathcal{B}\big(\Omega^{(2)}\big).
    \label{eqn:BW_subarray}
\end{align}
Remarkably, because \(\Omega^{(3)}\) is entirely redundant, removing the sub-array \(\mathcal{P}^{(3)}\) does not alter the total bandwidth \(\mathcal{B}(\Omega)\), meaning it does not reduce the available distance-domain DoF.

\begin{figure}[ht]
    \centering
    \includegraphics[width=9cm]{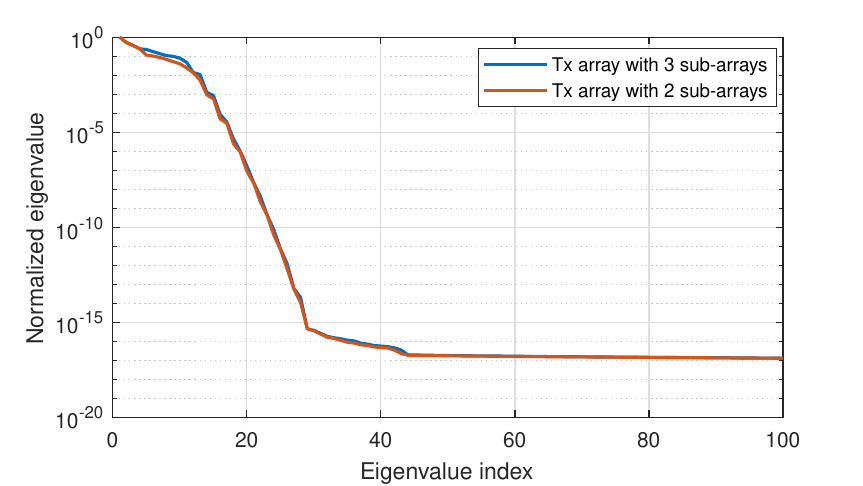}
    \caption{Eigenvalue distribution for the 3-subarray versus the 2-subarray configuration (final sub-array removed) from the example in Fig.~\ref{fig:modular_ULA_broadside_collinear_elements}.}
    \label{fig:two_modular_DoFcomparison}
\end{figure}

To verify this numerically, we simulate a setup with \([r_{\min},r_{\max}]=[100\lambda,\,200\lambda]\). The outer sub-arrays span \([80\lambda,\,100\lambda]\), and the inner sub-array spans \([0,\,40\lambda]\). As shown in Fig.~\ref{fig:two_modular_DoFcomparison}, the eigenvalue distributions for the full 3-subarray configuration and the truncated 2-subarray configuration (with \(\mathcal{P}^{(3)}\) removed) are essentially identical. Both strictly align with the analytical DoF of \(13\) predicted by \eqref{eqn:BW_subarray}.

\subsection{Distance-Domain DoF Comparison Between Modular Arrays and Single Array}

\subsubsection{Distance-domain DoF under physical length constraint}

\begin{figure}[t]
    \centering
    \includegraphics[width=9cm]{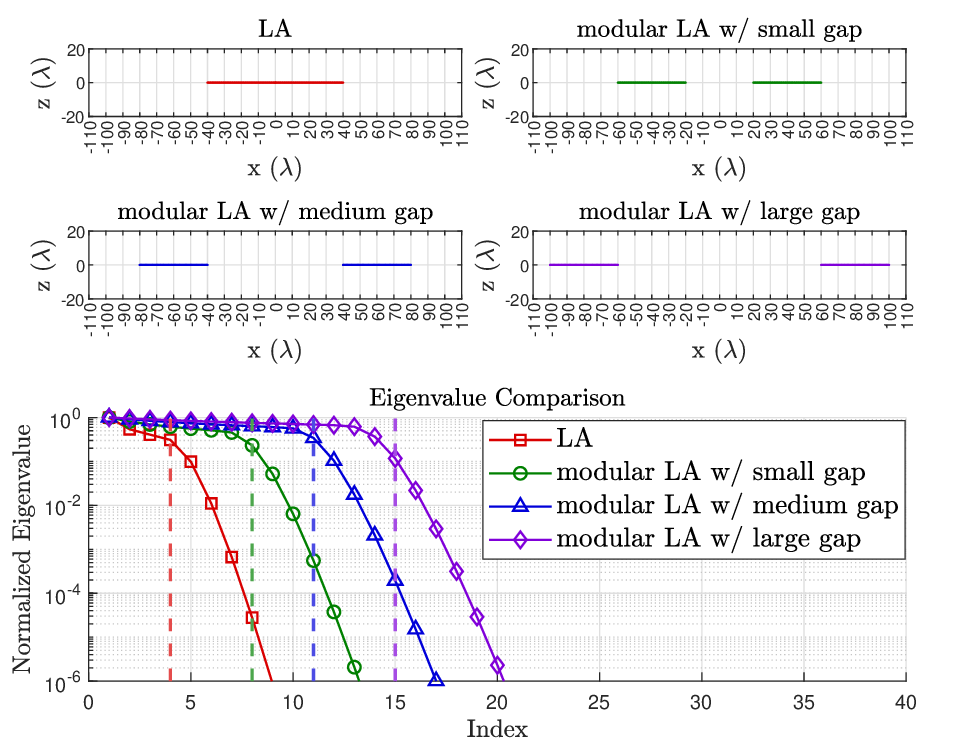}
    \caption{Element positions of four array configurations in the xz-plane (top panels) and the corresponding eigenvalue distributions (bottom panel) of the LoS channel between $\mathcal{P}$ and $\mathcal{Q}$. The Tx array $\mathcal{P}$ is either a single-piece linear array or a symmetric modular linear array. Both configurations share the same total active physical length $L$, and the Rx array $\mathcal{Q}$ spans from $r_{\min}=200\lambda$ to $r_{\max}=2000\lambda$. The vertical dotted lines represent the analytical DoF in (\ref{eqn:simplified_analytical_DoF}).}
    \label{fig:different_size_gap}
\end{figure}

Firstly, we compare a conventional single-piece linear array and a symmetric modular linear array, maintaining the same total active physical length \(L\). For a symmetric two-module array separated by a central gap of size \(G\), the sub-arrays span the intervals \([-(L+G)/2, -G/2]\) and \([G/2, (L+G)/2]\). Because the main-lobes of \( \Omega^{(1)} \) and \( \Omega^{(2)} \) are identical and perfectly overlap, only one sub-array determines the distance-domain DoF of the entire array, yielding:
\begin{align}
    \frac{L(L+2G)}{8\lambda}\left(r^{-1}_{\min}-r^{-1}_{\max}\right) + O(1).
    \label{eqn:DoF_gap}
\end{align}
Equation (\ref{eqn:DoF_gap}) shows that for a fixed active array length \(L\), increasing the gap size \(G\) strictly expands the distance-domain DoF.

\begin{figure}[ht]
    \centering
    \includegraphics[width=5cm]{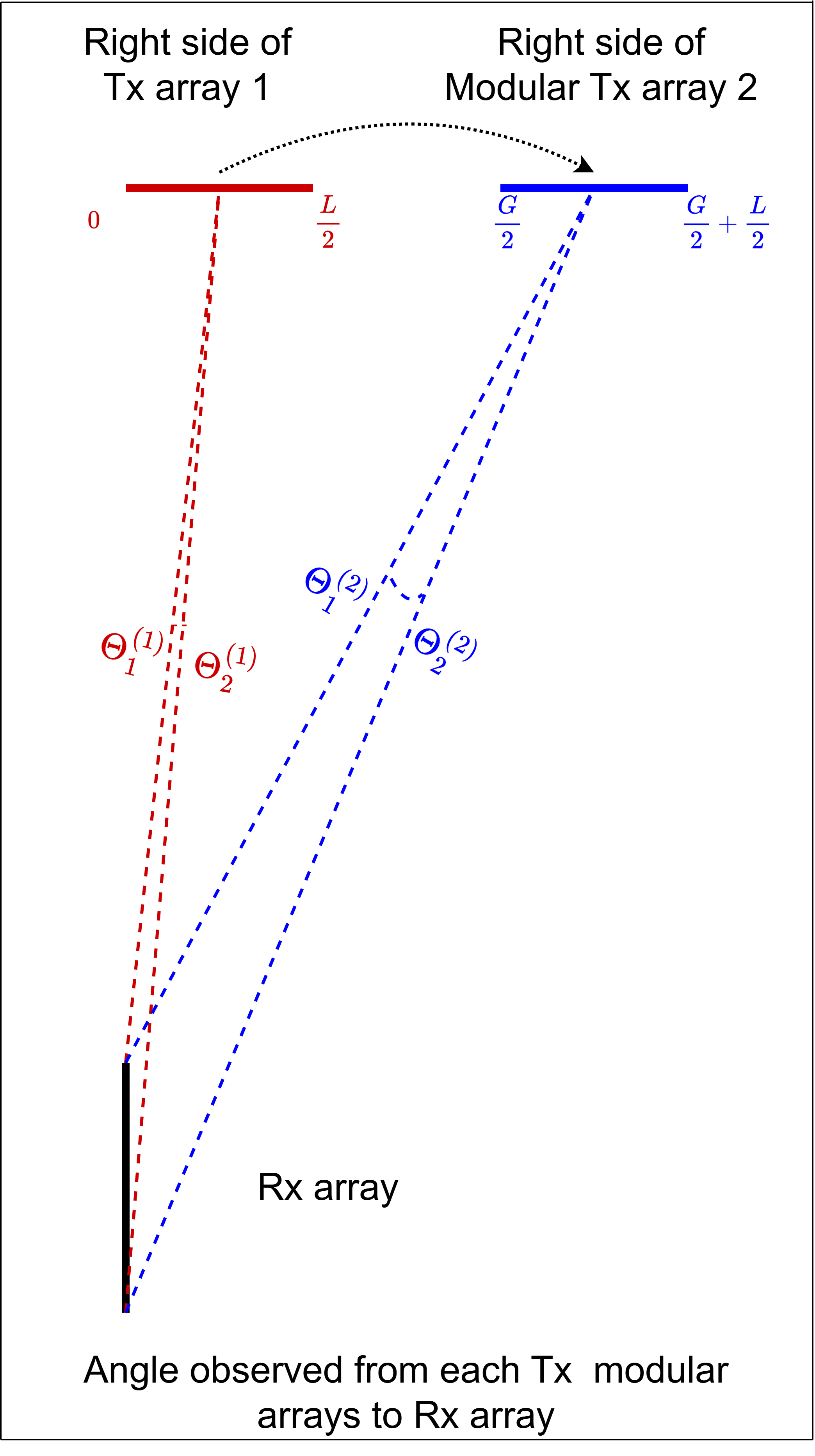}
    \caption{Geometric comparison of the subtended angles between a single-piece Tx array (left) and a modular Tx array with a central gap (right). Due to structural symmetry, only the right half of each Tx configuration is depicted.}
    \label{fig:modular_physical_explanation}
\end{figure}

\textcolor{black}{To intuitively explain this gain, we apply the signal space approach \cite{poon2005degrees}, where spatial DoF is proportional to the active array length $L$ and the difference in directional cosines of the receiver boundaries: $\text{DoF} \propto L (\cos\Theta_1 - \cos\Theta_2) = 2L \sin(\frac{\Theta_2 - \Theta_1}{2}) \sin(\frac{\Theta_2 + \Theta_1}{2})$. Here, $\Theta_1$ and $\Theta_2$ denote the extreme angles of the Rx array observed from the Tx elements.}

\textcolor{black}{As illustrated in Fig.~\ref{fig:modular_physical_explanation}, laterally shifting the active elements outward by introducing a gap $G$ induces two competing geometric effects:
\begin{itemize}
    \item \textbf{Negligible decrease in $\sin(\frac{\Theta_2 + \Theta_1}{2})$:} Since the near-field Rx array distances are much greater than the lateral shift, the angles remain close to $\pi/2$. Consequently, this sine term is nearly $1$, and the lateral shift barely reduces it.
    \item \textbf{Significant increase in $\sin(\frac{\Theta_2 - \Theta_1}{2})$:} Moving the Tx elements away from the collinear axis dramatically expands the angular spread $(\Theta_2 - \Theta_1)$ subtended by the deep longitudinal Rx array.
\end{itemize}
Since the substantial growth of the difference term far outpaces the marginal reduction in the sum term, the overall distance-domain DoF strictly increases.} This theoretical gain is confirmed numerically in Fig.~\ref{fig:different_size_gap}, demonstrating that a modular array can exploit additional spatial resources compared to a single-piece array of identical active physical length.

\subsubsection{Spatial DoF under extreme edge constraint}
\begin{figure}[ht]
    \centering
    \includegraphics[width=7cm]{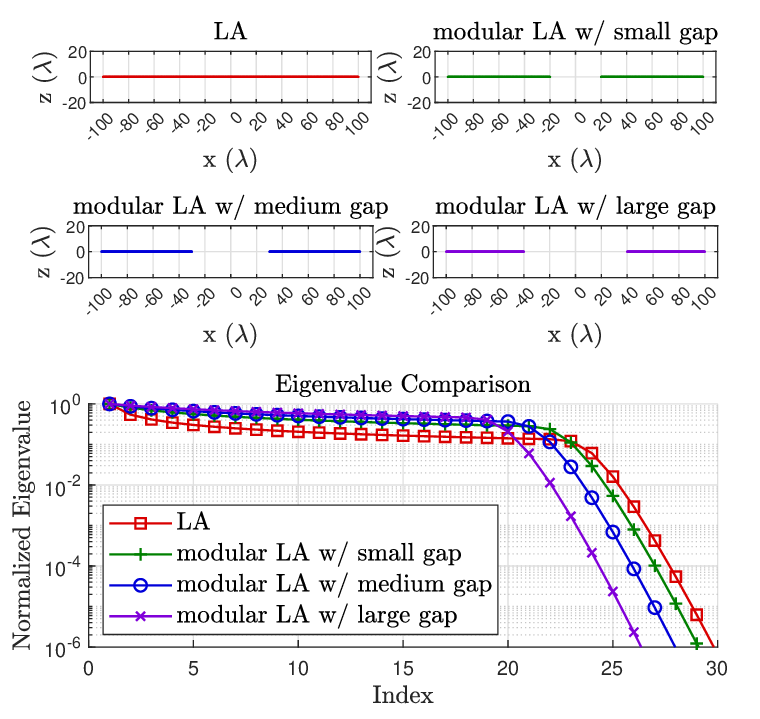}
    \caption{Element positions of four array configurations in the xz-plane (top panels) and the corresponding eigenvalue distributions (bottom panel) of the LoS channel where the Tx array is either a single-piece linear array or a symmetric modular linear array. Both configurations are constrained within the same extreme outer edges, and the Rx array spans from $r_{\min}=200\lambda$ to $r_{\max}=2000\lambda$.}
    \label{fig:different_hole_same_outer_size}
\end{figure}

Furthermore, consider a scenario where both a single-piece Tx array and a modular Tx array are strictly bounded within the extreme edges \([-L/2, L/2]\) (representing a maximum physical footprint of \(L\)). For the modular array, a central hole of length \(\alpha L\) (\(0<\alpha<1\)) is introduced, reducing the effective active length to \((1-\alpha)L\). Under these strict boundary constraints, a centered gap maximizes the distance-domain DoF, evaluating to:
\begin{align}
    \frac{L^2(1-\alpha^2)}{8\lambda}\left(r^{-1}_{\min}-r^{-1}_{\max}\right) + O(1).
\end{align}
Thus, although the modular array's effective active length is reduced by a factor of \(1-\alpha\), the resulting loss in spatial DoF is only of order \(\alpha^2\). For example, as seen in Fig.~\ref{fig:different_hole_same_outer_size}, a modular array with a \(40\%\) reduction in active length (\(\alpha = 0.4\)) experiences only a roughly \(16\%\) loss in spatial DoF. This confirms that the central region of the Tx array contributes minimally to the distance-domain DoF, highlighting a key advantage of modular arrays: they achieve comparable spatial multiplexing capabilities while significantly reducing the required hardware area.

%% file: discussion.tex
\textcolor{black}{\section{Discussion: Limitations and Future Directions}
While the proposed analytical framework provides a closed-form expression and valuable geometric insights into the distance-domain DoF, it is based on several idealized assumptions. These limitations inherently define the scope of our current contribution and establish a clear roadmap for future extensions:}

\begin{itemize}
    \item \textcolor{black}{\textbf{Continuous Aperture Idealization vs. Discrete Arrays:} This model assumes an infinite number of elements with infinitesimal element spacing to derive an upper bound for the distance-domain DoF. While mathematically convenient and insightful, this idealized framework is not directly implementable. To practically realize the distance-domain DoF, specific sampling strategies for both the transmitters and receivers are required, which will be addressed in our future work in \cite{duong2025sampling}.}
    \item \textcolor{black}{\textbf{LoS-Dominant Environments vs. Multipath Scattering:} The analysis is strictly restricted to the Line-of-Sight (LoS) component. While LoS models are highly valid in sparse-scattering environments such as mmWave and THz frequency bands, they do not capture the full channel complexity of rich scattering environments. Future work can incorporate Non-Line-of-Sight (NLoS) components to evaluate wireless channels with rich scattering, which typically occur at lower frequency carriers.}
    \item \textcolor{black}{\textbf{1D Observation Region vs. 3D observation region:} In this current work, we assume a 1D observation region to effectively isolate the angular-domain DoF and mathematically extract the distance-domain DoF. Such geometry corresponds to collinear users, which represents a highly specific multi-user topology. Future research can generalize this framework to 2D or 3D user distributions, enabling the joint optimization of distance- and angular-domain multiplexing.}
\end{itemize}

%% file: appendix_v4.tex

\section*{Appendix A: Proof of Lemma~\lowercase{\ref{lem:kernel_transformation}}}


Assume that $e_k(r)$ and $\lambda_k$ are an eigenfunction and eigenvalue of the operator ${\mathcal{V}}$ with kernel ${v}(r,r')$, so that
\begin{align}
	\lambda_k \, e_k(r) = \int_{r_{\min}}^{r_{\max}} {v}(r,r')\, e_k(r')\, dr, \quad r \in [r_{\min},r_{\max}].
	\label{eq:original_eigen_eq}
\end{align}
By performing the variable substitution $ \zeta = 1/r$ with the factor $dr =  d\zeta/ \zeta^2$) in \eqref{eq:original_eigen_eq}, we obtain
\begin{align*}
	\lambda_k\, e_k\!\left(\frac{1}{ \zeta}\right)
	& = \int_{r^{-1}_{\max}}^{r^{-1}_{\min}} {v}\!\left(\frac{1}{ \zeta}, \frac{1}{ \zeta'}\right)\, e_k\!\left(\frac{1}{ \zeta'}\right) \left(\frac{1}{ \zeta'^2}\,  d\zeta'\right).
\end{align*}
Multiplying both sides by $1/ \zeta $ and let $f_k( \zeta ) = e_k(1/ \zeta )/ \zeta $:
\begin{align}
	\lambda_k\, f_k( \zeta )
	& = \frac{1}{ \zeta }\, \lambda_k\, e_k\!\left(\frac{1}{ \zeta }\right) \nonumber \\
	& = \frac{1}{ \zeta } \int_{r^{-1}_{\max}}^{r^{-1}_{\min}} {v}\!\left(\frac{1}{ \zeta }, \frac{1}{ \zeta'}\right)\, e_k\!\left(\frac{1}{ \zeta'}\right) \left(\frac{1}{ \zeta'^2}\,  d\zeta'\right) \nonumber \\
	& = \int_{r^{-1}_{\max}}^{r^{-1}_{\min}} \left[\frac{1}{ \zeta\,  \zeta'}\, {v}\!\left(\frac{1}{ \zeta}, \frac{1}{ \zeta'}\right)\right] \left[ \frac{1}{ \zeta'} e_k\!\left(\frac{1}{ \zeta'}\right) \right]\,  d\zeta' \nonumber \\
	& = \int_{r^{-1}_{\max}}^{r^{-1}_{\min}} g( \zeta, \zeta')\, f_k( \zeta')\,  d\zeta',
	\label{eq:transformed_eigen_eq}
\end{align}
where we define the transformed kernel as
\begin{align}
	g( \zeta, \zeta') = \frac{1}{ \zeta\,  \zeta'}\, {v}\!\left(\frac{1}{ \zeta}, \frac{1}{ \zeta'}\right).
	\label{eq:transformed_kernel}
\end{align}
Thus, $f_k( \zeta)$ and $\lambda_k$ satisfy the eigenvalue equation for the operator $\mathcal{G}$ with kernel $g( \zeta, \zeta')$, which completes the proof.

\section*{Appendix B: Proof of Lemma~\lowercase{\ref{lem:fourier_closedform}}}


We use the \emph{scaled Fourier transform} defined as
\[
\tilde{g}(\xi)=\int_{-\infty}^{\infty} g( \zeta )e^{-j2\pi \xi \frac{ \zeta }{L_\zeta}}d \zeta ,
\]
where \( L_\zeta=r^{-1}_{\min}-r^{-1}_{\max} \). We express \( g( \zeta )= g_0( \zeta ) w( \zeta ) \) where \(w( \zeta ) = \mathbf{1}_{\{| \zeta|\leq L_\zeta\}}\) is the window function with duration \(2T\) and \( g_0( \zeta ) \) is given by:
\[
g_0( \zeta )=\iint_{\mathbf{p}\in\mathscr{P}} \exp\!\Bigl(-j\frac{\pi}{\lambda}\|\mathbf{p}\|^2\, \zeta \Bigr)d\mathbf{p}.
\]
while the Fourier transform of \(w( \zeta )\) is simply \( \tilde{w}(\xi)= 2\,\mathrm{sinc}(2\xi)\) with \(\mathrm{sinc}(x)=\frac{\sin(\pi x)}{\pi x}\). In addition, interchanging the order of integration shows that the scaled Fourier transform of \(g_0( \zeta )\) is
\begin{align}
    \tilde{g}_0(\xi) & = \int_{-\infty}^{\infty} \left[ \iint_{\mathbf{p}\in\mathscr{P}} e^{-j\frac{\pi}{\lambda}\|\mathbf{p}\|^2\, \zeta } d\mathbf{p} \right] e^{-j2\pi \xi \frac{ \zeta }{L_\zeta}}d \zeta  \nonumber \\
    & = \iint_{\mathbf{p}\in\mathscr{P}} \left[ \int_{-\infty}^{\infty} e^{-j2\pi \left(\frac{\|\mathbf{p}\|^2}{2\lambda} + \frac{\xi}{L_\zeta} \right)\, \zeta } d \zeta  \right] d\mathbf{p} \nonumber \\
    & = L_\zeta\iint_{\mathbf{p} \in \mathscr{P}} \delta\!\Bigl(\xi + \frac{\|\mathbf{p}\|^2T}{2\lambda}\Bigr)d\mathbf{p}
\end{align}
By the convolution theorem, the Fourier transform of \(g( \zeta)\) is the convolution of the Fourier transforms of the two factors. And by ignoring the constant scalar coefficients, we obtain:
\[
\tilde{g}(\xi) = \tilde{g}_0(\xi)*\mathrm{sinc}(2\xi),
\]